\documentclass[amsmath,aps,showpacs,a4paper,10pt]{revtex4}

 \usepackage{epsf}
 \usepackage{graphicx}    
 \usepackage{longtable}   

\begin{document}

 \newcommand{\be}[1]{\begin{equation}\label{#1}}
 \newcommand{\ee}{\end{equation}}
 \newcommand{\bea}{\begin{eqnarray}}
 \newcommand{\eea}{\end{eqnarray}}
 \def\disp{\displaystyle}

 \def\gsim{ \lower .75ex \hbox{$\sim$} \llap{\raise .27ex \hbox{$>$}} }
 \def\lsim{ \lower .75ex \hbox{$\sim$} \llap{\raise .27ex \hbox{$<$}} }

 \begin{titlepage}

 \begin{flushright}
 arXiv:1410.3960
 \end{flushright}

 \title{\Large \bf Cosmological Models and Gamma-Ray~Bursts
 Calibrated~by~Using~Pad\'e~Method}

 \author{Jing~Liu\,}
 \email[\,email address:\ ]{liujing5972@163.com}
 \affiliation{School of Physics,
 Beijing Institute of Technology, Beijing 100081, China}

 \author{Hao~Wei\,}
 \thanks{\,Corresponding author}
 \email[\,email address:\ ]{haowei@bit.edu.cn}
 \affiliation{School of Physics,
 Beijing Institute of Technology, Beijing 100081, China}

 \begin{abstract}\vspace{1cm}
 \centerline{\bf ABSTRACT}\vspace{2mm}
 Gamma-ray bursts (GRBs) are among the most powerful sources
 in the universe. In the recent years, GRBs have been proposed
 as a complementary probe to type Ia supernovae (SNIa).
 However, as is well known, there is a circularity problem
 in the use of GRBs to study cosmology. In this work, based on
 the Pad\'e approximant, we propose a new cosmology-independent
 method to calibrate GRBs. We consider a sample consisting of
 138 long Swift GRBs and obtain 79 calibrated long GRBs at
 high-redshift $z>1.4$ (named Mayflower sample) which can be used to
 constrain cosmological models without the circularity problem.
 Then, we consider the constraints on several cosmological models
 with these 79 calibrated GRBs and other observational data.
 We show that GRBs are competent to be a complementary probe to
 the other well-established cosmological observations.
 \end{abstract}


 \maketitle

 \end{titlepage}

 \renewcommand{\baselinestretch}{1.0}


\section{Introduction}\label{sec1}

Based on the observations of type Ia supernovae (SNIa), the
 current acceleration of the universe was firstly discovered in
 1998~\cite{r1}. This great discovery hints the existence
 of a new component with negative pressure called dark energy,
 or a modification to general relativity on the cosmological
 scale. In order to understand the cosmic acceleration,
 astronomers have made much effort on the cosmological
 observations. Besides SNIa, as is well known, there are other
 well-established cosmological observations such as baryon
 acoustic oscillations (BAO) and cosmic microwave background (CMB).

In the recent years, Gamma-Ray Bursts (GRBs)~\cite{r2,r3,r4,r5,r55}
 have been proposed as a complementary probe to SNIa. Their
 high energy photons in the gamma-ray band are almost immune
 to dust extinction, and hence they have been observed up to
 redshift $z \sim 8-9$~\cite{r6,r7}, well beyond the observed
 redshift range of SNIa, namely $z<2$~\cite{r8}. Thus, we might
 use GRBs to explore the early universe in the high redshift
 range which is difficult to access by other cosmological
 probes. To our knowledge, using GRBs to constrain the cosmological
 models was firstly performed by Dai {\it et al.}~\cite{r9}.
 However, there is a so-called ``circularity problem''~\cite{r2} in
 the direct use of GRBs, mainly due to the lack of a set of
 low-redshift GRBs at $z<0.1$ which are cosmology-independent.
 To calibrate the empirical GRB luminosity relations, one
 should assume a particular cosmological model with some model
 parameters {\it a priori}. Therefore, when one uses these
 ``calibrated'' GRBs (which are actually cosmology-dependent)
 to constrain cosmological models, the circularity problem appears.
 To alleviate the circularity problem, some statistical methods
 were proposed, including the scatter method~\cite{r10}, the
 luminosity distance method~\cite{r10,r11}, the Bayesian
 method~\cite{r12}, and so on. However, they still cannot solve
 the circularity problem completely.

Up to date, in the literature there are several
 cosmology-independent methods to avoid the circularity
 problem. For example, Li {\it et al.}~\cite{r13} proposed that
 one can treat the parameters involved in the empirical GRB
 correlation relation as free parameters, and determine them
 simultaneously with the cosmological model parameters by using
 GRBs data together with other observational data. However, for
 any given cosmological model, this method can always obtain
 some parameters for the cosmological model and the empirical
 GRB luminosity relation. In this sense, any cosmological model
 is ``viable'' (except for a few obviously absurd models), and
 hence this method cannot be used to rule out any cosmological
 model. So, it is not a satisfactory method to solve the
 circularity problem completely. A completely
 cosmology-independent method was proposed by
 Liang {\it et al.}~\cite{r14}. The key idea is using distance
 ladder to calibrate GRBs. Similar to calibrating SNIa as
 secondary standard candles by using Cepheid variables which
 are primary standard candles, we can also calibrate GRBs as
 standard candles with a large amount of SNIa.
 Liang {\it et al.}~\cite{r14} proposed to divide GRBs into
 two groups, whose redshifts are $z<1.4$ and $z\geq 1.4$,
 respectively. Using a cubic interpolation method, one can
 obtain the distance modulus of a GRB at a given low-redshift
 $z<1.4$ by interpolating from the Hubble diagrim of SNIa.
 Since the distance moduli of SNIa are obtained directly from
 observations, this method is completely cosmology-independent.
 We can calibrate the empirical GRB luminosity relations with
 these low-redshift GRBs at $z<1.4$, and then derive the
 distance moduli of the high-redshift GRBs at $z\geq 1.4$ by
 using the calibrated empirical GRB luminosity relations.
 Obviously, the calibrated high-redshift GRBs can be used to
 constrain cosmological models without the circularity problem.
 In~\cite{r15,r16}, Wei {\it et al.} have further developed
 this method. They considered a sample of 109 GRBs, and
 obtained 59 calibrated high-redshift GRBs (named Hymnium
 sample)~\cite{r16} which can be used to constrain cosmological
 models. It is worth noting that almost at the same time,
 Kodama {\it et al.}~\cite{r17} proposed a similar method
 using also the idea of distance ladder. Instead of the cubic
 interpolation method used in~\cite{r14,r15,r16}, Kodama
 {\it et al.}~\cite{r17} found an completely
 empirical formula for the luminosity distance of SNIa at
 redshift $0.359<z<1.755$, namely
 \be{eq1}
 \frac{d_L}{10^{27}\, \rm cm}=
 14.57\times z^{1.02}+7.16\times z^{1.76}\,.
 \ee
 Then, similar to~\cite{r14,r15,r16}, Kodama {\it et al.}~\cite{r17}
 calibrated the empirical GRB luminosity relations with these
 low-redshift GRBs at $z\leq 1.755$, and then derive the
 distance moduli of the high-redshift GRBs at $z>1.755$ by
 using the calibrated empirical GRB luminosity relations.
 Obviously, this method has a fatal drawback. As is also
 admitted by them, the empirical formula in Eq.~(\ref{eq1}) was
 written purely by hand without any theoretical foundation.
 Thus, this method has not been widely used in the literature.
 On the other hand, Capozziello {\it et al.}~\cite{r18}
 considered a cosmography method (see also e.g.~\cite{r19}).
 They expanded the luminosity distance $d_L$ by using the
 Taylor series up to high order in redshift $z$,
 whose coefficients are characterized by the cosmographic
 parameters~\cite{r56}, namely the Hubble constant $H_0$,
 deceleration parameter $q_0$, jerk $j_0$, snap $s_0$ and
 lerk $l_0$. Then, they fitted this luminosity distance $d_L$
 to SNIa dataset, and obtained the best-fit cosmographic
 parameters with $1\sigma$ uncertainty. So, the luminosity
 distance of GRBs can be derived from the $d_L$ expansion with
 the cosmographic parameters calibrated by SNIa. Obviously,
 this method is also cosmology-independent. However, it is
 well known that the Taylor series converges only for small
 $z$, and it might diverge at higher redshifts (especially
 when $z\,\gsim\,1$). This shortcoming cannot be completely
 cured by replacing the redshift $z$ with the so-called
 $y$-shift $y\equiv z/(1+z)$, because the error of Taylor
 approximation throwing away the higher order terms will
 become unacceptably large when $y$ is close to 1 (say, when
 $z>9$). Unlike the above methods, Wang~\cite{r20} considered
 the calibration of GRBs by using the data of GRBs internally,
 without invoking any external datasets (e.g. SNIa). Based on
 this method, considering the sample of 109 GRBs given by
 Wei~\cite{r16}, Xu~\cite{r21} derived five data points of
 distance measurements which do not depend on any cosmological
 models. However, the method of Wang~\cite{r20} is still under
 a slight suspicion that it is not so model-independent
 actually. In this method, when one determines the statistical
 errors of correlation parameters and the systematic error, a
 particular $\Lambda$CDM model with $\Omega_{m0}=0.27$ was
 assumed, although they claimed that the correlation parameters
 themselves do not depend on this assumed $\Lambda$CDM
 model~\cite{r20,r21}. On the other hand, in this method the
 absolute magnitude of GRBs is unknown, only the slopes of GRBs
 correlations can be used as cosmological constraints. As a
 result, it is shown in e.g.~\cite{r21} that the constraints on
 cosmological models using this method is looser than the one
 of~\cite{r16} which uses the method proposed by Liang {\it et
 al.}~\cite{r14}. In fact, several further potential drawbacks
 of the method of Wang~\cite{r20} were listed in, for example,
 the last section of~\cite{r21}.

In the present work, we try to further develop the
 cosmology-independent method to calibrate GRBs. Inspired by
 the methods proposed by Liang {\it et al.}~\cite{r14},
 Kodama {\it et al.}~\cite{r17}, and Capozziello
 {\it et al.}~\cite{r18}, we propose a new method engrafting
 the advantages of these three methods without their drawbacks.
 We keep the key idea of distance ladder and the main framework
 of the method used in e.g.~\cite{r14,r15,r16}, but change the
 method to obtain the distance moduli (or luminosity distances
 equivalently) of the low-redshift GRBs at $z<1.4$. Instead of
 the purely empirical formula for the luminosity distance of
 SNIa in Eq.~(\ref{eq1})~\cite{r17}, or the Taylor expansion
 of the luminosity distance of SNIa~\cite{r18}, we consider
 the Pad\'e approximant, which can be regarded as
 a generalization of Taylor polynomial~\cite{r22,r22inst}.
 In mathematics, a Pad\'e approximant is the best
 approximation of a function by a rational function of given
 order~\cite{r22}. In fact, the Pad\'e approximant often
 gives better approximation of the function than truncating
 its Taylor series, and it may still work where the Taylor
 series does not converge~\cite{r22}. For any function $f(x)$,
 its corresponding Pad\'e approximant of order $(m,\,n)$ is
 given by the rational function~\cite{r22,r22inst}
 \be{eq2}
 f(x)=\frac{\alpha_0+\alpha_1 x+\cdots+\alpha_m x^m}{1+
 \beta_1 x+\cdots+\beta_n x^n}\,,
 \ee
 where $m$ and $n$ are both non-negative integers;
 $\alpha_i$ and $\beta_i$ are all constants. Obviously, it
 reduces to the Taylor polynomial when all $\beta_i=0$. It is
 worth noting that if we express the luminosity distance of
 SNIa at low redshift $z<1.4$ with the Pad\'e approximant, it
 is well motivated from the theoretical point of view, unlike
 the empirical formula in Eq.~(\ref{eq1}) purely written by
 hand~\cite{r17}. As mentioned above, the Pad\'e approximant
 can also avoid the divergence of Taylor polynomial at high
 redshift, unlike the case of cosmography method~\cite{r18}.

In this work, we consider a sample consisting of 138 long Swift
 GRBs (see e.g.~\cite{r58} for Swift mission). It includes 109
 long GRBs adopted directly from~\cite{r16}, which have 50
 low-redshift GRBs at $z<1.4$ and 59 high-redshift GRBs at $z>1.4$.
 In addition, we adopt other 29 long GRBs from~\cite{r23},
 which include 9 low-redshift GRBs (050126A, 050223, 050803,
 060904B, 100621A, 100816A, 101219B, 070508, 100414A) and 20
 high-redshift GRBs (100814A, 110213A, 100906A, 081203A,
 100728B, 080804, 110205A, 070110, 060714, 060607A, 050908,
 061222B, 060906, 060605, 060210, 050505, 060223A, 060510B,
 060522, 050814). In total, we have 59 low-redshift GRBs at
 $z<1.4$ and 79 high-redshift GRBs at $z>1.4$.

The rest of this paper is organized as followings. In
 Sec.~\ref{sec2}, we calibrate 138 GRBs with Union2.1 SNIa
 dataset using the method of Pad\'e approximant. We obtain 79
 calibrated GRBs at high redshift $z>1.4$ (named Mayflower
 sample) which can be used to constrain cosmological models
 without the circularity problem. In Sec.~\ref{sec3}, we
 consider the constraints on several cosmological models
 with these 79 calibrated GRBs and other observational data.
 In Sec.~\ref{sec4}, the conclusion and discussions are given.



 \begin{table}[ptbh]
 \begin{center}
 \begin{tabular}{l|cccccccc}
 \hline\hline
 \ $(m,\,n)$ & $(1,\,1)$ & $(1,\,2)$ & $(1,\,3)$ & $(1,\,4)$
  & $(2,\,1)$ & $(2,\,2)$ & $(2,\,3)$ & $(2,\,4)$ \\[0.5mm] \hline
 \ $\chi^2_{min}$ & ~~1626.99~~ & ~738.582~ & ~608.276~ & ~576.214~
  & ~719.261~ & ~571.932~ & ~561.947~ & ~561.032~\\[0.5mm]\hline
 \ $k$ & 3 & 4 & 5 & 6 & 4 & 5 & 6 & 7 \\[0.5mm]\hline
 \ $\chi^2_{min}/dof$~~ & 2.8197 & 1.2823 & 1.0579 & 1.0039 & 1.2487
  & 0.9950& 0.979& 0.9791 \\[0.5mm]\hline
 \ $\Delta$BIC & 1046.06 & 164.015 & 40.072 & 14.373 & 144.694
  & 3.728 & 0.106 & 5.554 \\[0.5mm]\hline
 \ $\Delta$AIC & 1059.15 & 172.741 & 44.435 & 14.373 & 153.42
  & 8.091 & 0.106 & 1.191 \\[0.5mm]\hline
 \hline
 \ $(m,\,n)$ & $(3,\,1)$ & $(3,\,2)$ & $(3,\,3)$ & $(3,\,4)$
  & $(4,\,1)$ & $(4,\,2)$ & $(4,\,3)$ & $(4,\,4)$ \\[0.5mm] \hline
 \ $\chi^2_{min}$ & 601.157	& 561.841 & 560.891 & 560.82
  & 573.314 & 561.006 & 560.82 & 560.82 \\[0.5mm]\hline
 \ $k$ & 5 & 6 & 7 & 8 & 6 & 7 & 8 & 9 \\[0.5mm]\hline
 \ $\chi^2_{min}/dof$~~ & 1.0455 & 0.9788 & 0.9789 & 0.9805
  & 0.9988 & 0.9791 & 0.9805 & 0.9822 \\[0.5mm]\hline
 \ $\Delta$BIC & 32.953 & 0 & 5.413	& 11.705 & 11.473 & 5.528
  & 11.705 & 18.068 \\[0.5mm]\hline
 \ $\Delta$AIC & 37.316 & 0 & 1.05 & 2.979 & 11.473 & 1.165
  & 2.979 & 4.979 \\[0.5mm]
 \hline\hline
 \end{tabular}
 \end{center}
 \caption{\label{tabinst} Comparing various Pad\'e approximants
 up to order $(4,\,4)$. See the text for details.}
 \end{table}



\section{Calibrating GRBs with the method of Pad\'e approximant}\label{sec2}

\begin{center}
 \begin{figure}[htbp]
 \centering
 \includegraphics[width=0.5\textwidth]{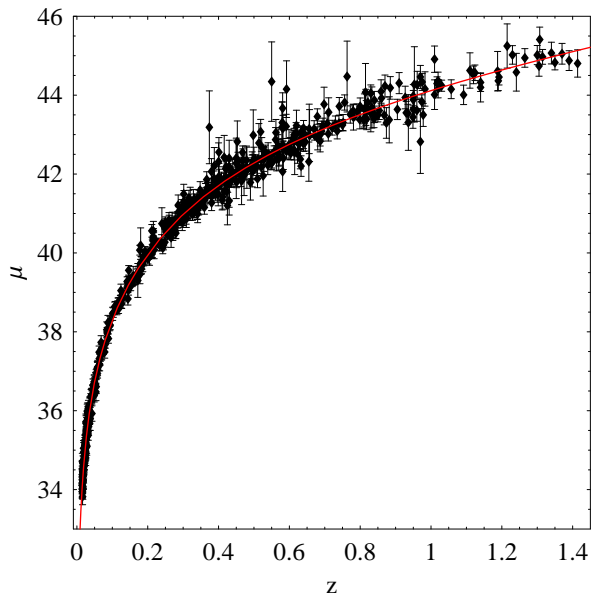}
 \caption{\label{fig1}
 The Hubble diagram of 580 Union2.1 SNIa  (black diamonds) and
 the distance moduli expressed in the Pad\'e approximant of
 order $(3,\,2)$ with the best-fit coefficients (red line).}
 \end{figure}
 \end{center}

In this section, we calibrate 138 GRBs with Union2.1 SNIa
 dataset~\cite{r24} (which consists of 580 SNIa) using the
 method of Pad\'e approximant. The first step is to find a
 formula for the distance moduli (or luminosity distances
 equivalently) of these 580 Union2.1 SNIa. Instead of purely
 empirical formula~\cite{r17} or Taylor expansion~\cite{r18},
 we consider the Pad\'e approximant given in Eq.~(\ref{eq2})
 which is well motivated from the theoretical point of view
 as mentioned above. Now, the question is how to choose the
 order $(m,\,n)$ of Pad\'e approximant. If the order is too
 low, the error of Pad\'e approximant will be unacceptably
 large. If the order is too high, the number of
 free coefficients are too much and the uncertainties will
 be large. To find the suitable order $(m,\,n)$, we test all
 the corresponding Pad\'e approximants up to order $(4,\,4)$
 one by one. For each Pad\'e approximant of given order, we
 fit the distance moduli in the expression of
 Pad\'e approximant to the real Union2.1 SNIa dataset, and
 minimize the corresponding $\chi^2$, namely
 \be{eq3}
 \chi^2=\sum_{i}\frac{\left[\mu_{obs}(z_i)
 -\mu_{pade}(z_i)\right]^2}{\sigma^2_i}\,,
 \ee
 where $\sigma$ is the corresponding $1\sigma$ error. Then,
 we compare all these 16 Pad\'e approximants up to order
 $(4,\,4)$. A conventional criterion for comparison in the
 literature is $\chi^2_{min}/dof$, in which the degree of
 freedom $dof=N-k$, whereas $N$ and $k$ are the number of
 data points and the number of free parameters, respectively.
 In addition, we also consider other two criterions used
 extensively in the literature, namely the so-called Bayesian
 Information Criterion (BIC) and Akaike Information Criterion
 (AIC). The BIC is defined by~\cite{r51}
 \be{eq3inst1}
 {\rm BIC}=-2\ln{\cal L}_{max}+k\ln N\,,
 \ee
 where ${\cal L}_{max}$ is the maximum likelihood.
 In the Gaussian cases, $\chi^2_{min}=-2\ln{\cal L}_{max}$.
 So, the difference in BIC between two fits is given by
 $\Delta{\rm BIC}=\Delta\chi^2_{min}+\Delta k \ln N$. The
 AIC is defined by~\cite{r52}
 \be{eq3inst2}
 {\rm AIC}=-2\ln{\cal L}_{max}+2k\,.
 \ee
%
%
 Accordingly, the difference in AIC between two fits is
 given by $\Delta{\rm AIC}=\Delta\chi^2_{min}+2\Delta k$.
 We present the results in Table~\ref{tabinst}. Notice that
 the Pad\'e approximant of order $(3,\,2)$ has been chosen
 to be the fiducial one when we calculate $\Delta$BIC and
 $\Delta$AIC. From Table~\ref{tabinst}, it is easy to see
 that the Pad\'e approximant of order $(3,\,2)$ is the best.
 So, we express the distance moduli in Pad\'e approximant
 of order $(3,\,2)$, namely
 \be{eq4}
 \mu_{pade}(z)=\frac{\alpha_0+\alpha_1 z+\alpha_2 z^2+
 \alpha_3 z^3}{1+\beta_1 z+\beta_2 z^2}\,.
 \ee
 We fit this formula for the distance moduli to the real 580
 Union2.1 SNIa dataset. By minimizing the corresponding
 $\chi^2$ in Eq.~(\ref{eq3}), we find that the best-fit
 coefficients (with $1\sigma$ errors) are given by
 $\alpha_0=30.1297\pm 0.9086$, $\alpha_1=2654.46\pm 898.233$,
 $\alpha_2=10740.4\pm 6173.28$, $\alpha_3=351.476\pm 262.753$,
 $\beta_1=71.3026\pm 25.3371$, $\beta_2=239.978\pm 140.868$,
 while $\chi^2_{min} = 561.841$. The corresponding covariance
 matrix~\cite{r25} (see also e.g.~\cite{r26,r27}) is given by
 \be{eq5}
 {\scriptsize
 \left(
 \begin{array}{cccccc}
   0.825595685565 & -802.86374851 &  -5283.1838241  &  -212.71322979  &  -22.600394329  &  -120.172829002 \\
   -802.86374851 &  806822.199728  &  5458365.5695  &  223635.10205  &   22757.10276006  &  124325.358406 \\
   -5283.1838241  &  5458365.5695  &  38109346.82922  &  1600878.89393  &  154260.002287  &  869566.838689 \\
   -212.71322979  &  223635.10205   &  1600878.89393  &  69039.099861 &  6328.7278164  &  36588.879649 \\
   -22.600394329  &  22757.10276006  &  154260.002287  &  6328.7278164  &  641.968817  &  3513.94073660 \\
   -120.172829002  &  124325.358406  &  869566.838689  &  36588.879649  &  3513.94073660  &  19843.704035 \\
 \end{array}
 \right)}
 \ee
 In Fig.~\ref{fig1}, we present the Hubble diagram of 580
 Union2.1 SNIa and the distance moduli expressed in the Pad\'e
 approximant of order $(3,\,2)$ with the best-fit coefficients.


 \begin{center}
 \begin{figure}[htbp]
 \centering
 \vspace{-3mm} 
 \includegraphics[width=0.98\textwidth]{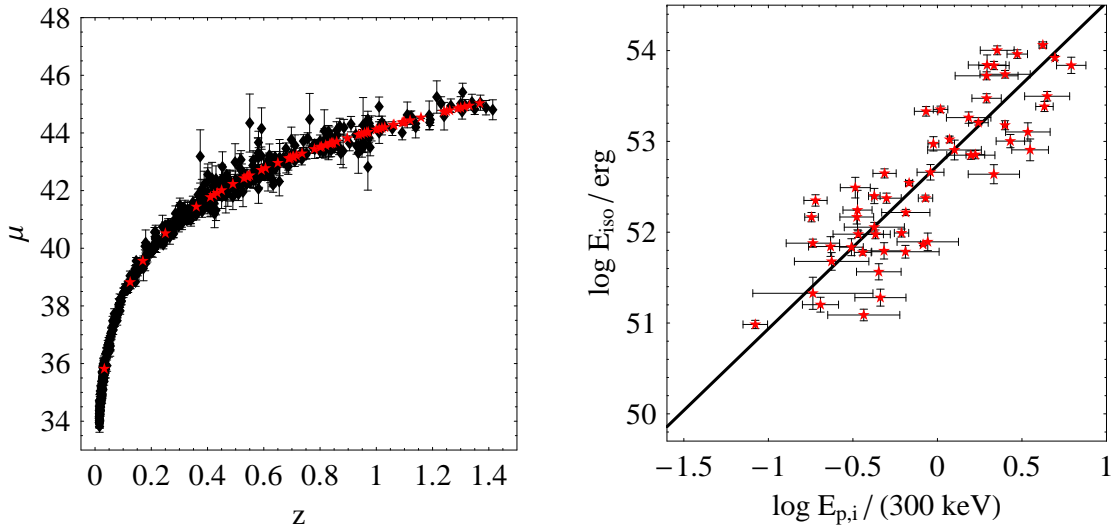}
 \caption{\label{fig1inst}
 Left panel: The Hubble diagram of 580 Union2.1 SNIa (black diamonds)
 and 59 low-redshift GRBs (red stars) whose distance moduli and
 errors are derived by using Pad\'e method. Noting that the
 error bars of GRBs are too small, one cannot clearly see
 them in this panel because they are hidden by the red stars of GRBs.
 Right panel: 59 low-redshift GRBs data (red stars) in the
 $\log E_{\rm p,i}\,/({\rm 300\,keV})-\log E_{\rm iso}\,/{\rm erg}$
 plane. The best-fit calibration line is also plotted. See the
 text for details.}
 \end{figure}
 \end{center}




 \begingroup
 \renewcommand{\baselinestretch}{0.6}
 \squeezetable
 \begin{table}[ptbh]
 \begin{center}
 \begin{tabular}{llccc}
 \hline\hline \ GRB & ~~~~~$z$ & ~~~~$S_{\rm bolo}~(10^{-5}~\rm erg\, cm^{-2})$
 & ~~~~~$E_{\rm p,i}~(\rm keV)$ & \ \ $\mu$ \\[1mm] \hline
 \ 060218	 & ~~~~ 0.0331	 & ~~~~ 2.2$\,\pm\,$0.1	 & ~~~~ 4.9$\,\pm\,$0.3	 & ~~~~ 35.819$\,\pm\,$0.017 \ \\
 \ 060614	 & ~~~~ 0.125	 & ~~~~ 5.9$\,\pm\,$2.4	 & ~~~~ 55.0$\,\pm\,$45.0	 & ~~~~ 38.825$\,\pm\,$0.014 \ \\
 \ 030329	 & ~~~~ 0.17	 & ~~~~ 21.5$\,\pm\,$3.8	 & ~~~~ 100.0$\,\pm\,$23.0	 & ~~~~ 39.563$\,\pm\,$0.013 \ \\
 \ 020903	 & ~~~~ 0.25	 & ~~~~ 0.016$\,\pm\,$0.004	 & ~~~~ 3.37$\,\pm\,$1.79	 & ~~~~ 40.519$\,\pm\,$0.013 \ \\
 \ 011121	 & ~~~~ 0.36	 & ~~~~ 24.3$\,\pm\,$6.7	 & ~~~~ 1060.0$\,\pm\,$265.0	 & ~~~~ 41.442$\,\pm\,$0.013 \ \\
 \ 020819B	 & ~~~~ 0.41	 & ~~~~ 1.6$\,\pm\,$0.4	 & ~~~~ 70.0$\,\pm\,$21.0	 & ~~~~ 41.774$\,\pm\,$0.013 \ \\
 \ 050803	 & ~~~~ 0.422	 & ~~~~ 0.41$\,\pm\,$0.09	 & ~~~~ 138.0$\,\pm\,$48.0	 & ~~~~ 41.847$\,\pm\,$0.013 \ \\
 \ 990712	 & ~~~~ 0.434	 & ~~~~ 1.4$\,\pm\,$0.3	 & ~~~~ 93.0$\,\pm\,$15.0	 & ~~~~ 41.919$\,\pm\,$0.013 \ \\
 \ 010921	 & ~~~~ 0.45	 & ~~~~ 1.8$\,\pm\,$0.2	 & ~~~~ 129.0$\,\pm\,$26.0	 & ~~~~ 42.012$\,\pm\,$0.013 \ \\
 \ 091127	 & ~~~~ 0.49	 & ~~~~ 2.34$\,\pm\,$0.28	 & ~~~~ 54.0$\,\pm\,$5.0	 & ~~~~ 42.23$\,\pm\,$0.014 \ \\
 \ 081007	 & ~~~~ 0.5295	 & ~~~~ 0.22$\,\pm\,$0.04	 & ~~~~ 61.0$\,\pm\,$15.0	 & ~~~~ 42.43$\,\pm\,$0.015 \ \\
 \ 090618	 & ~~~~ 0.54	 & ~~~~ 28.09$\,\pm\,$3.37	 & ~~~~ 257.0$\,\pm\,$41.0	 & ~~~~ 42.48$\,\pm\,$0.016 \ \\
 \ 100621A	 & ~~~~ 0.542	 & ~~~~ 5.78$\,\pm\,$0.66	 & ~~~~ 146.0$\,\pm\,$23.1	 & ~~~~ 42.49$\,\pm\,$0.016 \ \\
 \ 090424	 & ~~~~ 0.544	 & ~~~~ 5.9$\,\pm\,$1.15	 & ~~~~ 273.0$\,\pm\,$50.0	 & ~~~~ 42.499$\,\pm\,$0.016 \ \\
 \ 101219B	 & ~~~~ 0.55	 & ~~~~ 0.76$\,\pm\,$0.05	 & ~~~~ 108.5$\,\pm\,$12.4	 & ~~~~ 42.528$\,\pm\,$0.016 \ \\
 \ 050223	 & ~~~~ 0.5915	 & ~~~~ 0.13$\,\pm\,$0.02	 & ~~~~ 110.0$\,\pm\,$54.0	 & ~~~~ 42.716$\,\pm\,$0.017 \ \\
 \ 050525A	 & ~~~~ 0.606	 & ~~~~ 2.6$\,\pm\,$0.5	 & ~~~~ 127.0$\,\pm\,$10.0	 & ~~~~ 42.779$\,\pm\,$0.017 \ \\
 \ 050416A	 & ~~~~ 0.65	 & ~~~~ 0.09$\,\pm\,$0.01	 & ~~~~ 25.1$\,\pm\,$4.2	 & ~~~~ 42.961$\,\pm\,$0.018 \ \\
 \ 080916	 & ~~~~ 0.689	 & ~~~~ 0.79$\,\pm\,$0.08	 & ~~~~ 184.0$\,\pm\,$18.0	 & ~~~~ 43.113$\,\pm\,$0.018 \ \\
 \ 020405	 & ~~~~ 0.69	 & ~~~~ 8.4$\,\pm\,$0.7	 & ~~~~ 354.0$\,\pm\,$10.0	 & ~~~~ 43.117$\,\pm\,$0.018 \ \\
 \ 970228	 & ~~~~ 0.695	 & ~~~~ 1.3$\,\pm\,$0.1	 & ~~~~ 195.0$\,\pm\,$64.0	 & ~~~~ 43.136$\,\pm\,$0.019 \ \\
 \ 060904B	 & ~~~~ 0.703	 & ~~~~ 0.28$\,\pm\,$0.06	 & ~~~~ 135.0$\,\pm\,$41.0	 & ~~~~ 43.166$\,\pm\,$0.019 \ \\
 \ 991208	 & ~~~~ 0.706	 & ~~~~ 17.2$\,\pm\,$1.4	 & ~~~~ 313.0$\,\pm\,$31.0	 & ~~~~ 43.177$\,\pm\,$0.019 \ \\
 \ 041006	 & ~~~~ 0.716	 & ~~~~ 2.3$\,\pm\,$0.6	 & ~~~~ 98.0$\,\pm\,$20.0	 & ~~~~ 43.214$\,\pm\,$0.019 \ \\
 \ 090328	 & ~~~~ 0.736	 & ~~~~ 8.93$\,\pm\,$2.06	 & ~~~~ 1028.0$\,\pm\,$312.0	 & ~~~~ 43.287$\,\pm\,$0.019 \ \\
 \ 030528	 & ~~~~ 0.78	 & ~~~~ 1.4$\,\pm\,$0.2	 & ~~~~ 57.0$\,\pm\,$9.0	 & ~~~~ 43.441$\,\pm\,$0.019 \ \\
 \ 051022	 & ~~~~ 0.8	 & ~~~~ 32.6$\,\pm\,$3.1	 & ~~~~ 754.0$\,\pm\,$258.0	 & ~~~~ 43.508$\,\pm\,$0.02 \ \\
 \ 100816A	 & ~~~~ 0.8049	 & ~~~~ 0.43$\,\pm\,$0.01	 & ~~~~ 246.73$\,\pm\,$8.48	 & ~~~~ 43.524$\,\pm\,$0.02 \ \\
 \ 070508	 & ~~~~ 0.82	 & ~~~~ 4.55$\,\pm\,$1.14	 & ~~~~ 378.56$\,\pm\,$138.32	 & ~~~~ 43.574$\,\pm\,$0.02 \ \\
 \ 970508	 & ~~~~ 0.835	 & ~~~~ 0.34$\,\pm\,$0.07	 & ~~~~ 145.0$\,\pm\,$43.0	 & ~~~~ 43.623$\,\pm\,$0.02 \ \\
 \ 060814	 & ~~~~ 0.84	 & ~~~~ 3.8$\,\pm\,$0.4	 & ~~~~ 473.0$\,\pm\,$155.0	 & ~~~~ 43.639$\,\pm\,$0.02 \ \\
 \ 990705	 & ~~~~ 0.842	 & ~~~~ 9.8$\,\pm\,$1.4	 & ~~~~ 459.0$\,\pm\,$139.0	 & ~~~~ 43.645$\,\pm\,$0.02 \ \\
 \ 000210	 & ~~~~ 0.846	 & ~~~~ 8.0$\,\pm\,$0.9	 & ~~~~ 753.0$\,\pm\,$26.0	 & ~~~~ 43.658$\,\pm\,$0.02 \ \\
 \ 040924	 & ~~~~ 0.859	 & ~~~~ 0.49$\,\pm\,$0.04	 & ~~~~ 102.0$\,\pm\,$35.0	 & ~~~~ 43.699$\,\pm\,$0.02 \ \\
 \ 091003	 & ~~~~ 0.8969	 & ~~~~ 4.75$\,\pm\,$0.79	 & ~~~~ 810.0$\,\pm\,$157.0	 & ~~~~ 43.816$\,\pm\,$0.021 \ \\
 \ 080319B	 & ~~~~ 0.937	 & ~~~~ 49.7$\,\pm\,$3.8	 & ~~~~ 1261.0$\,\pm\,$65.0	 & ~~~~ 43.936$\,\pm\,$0.022 \ \\
 \ 071010B	 & ~~~~ 0.947	 & ~~~~ 0.74$\,\pm\,$0.37	 & ~~~~ 101.0$\,\pm\,$20.0	 & ~~~~ 43.965$\,\pm\,$0.023 \ \\
 \ 970828	 & ~~~~ 0.958	 & ~~~~ 12.3$\,\pm\,$1.4	 & ~~~~ 586.0$\,\pm\,$117.0	 & ~~~~ 43.997$\,\pm\,$0.023 \ \\
 \ 980703	 & ~~~~ 0.966	 & ~~~~ 2.9$\,\pm\,$0.3	 & ~~~~ 503.0$\,\pm\,$64.0	 & ~~~~ 44.02$\,\pm\,$0.024 \ \\
 \ 091018	 & ~~~~ 0.971	 & ~~~~ 0.3$\,\pm\,$0.03	 & ~~~~ 55.0$\,\pm\,$20.0	 & ~~~~ 44.034$\,\pm\,$0.024 \ \\
 \ 980326	 & ~~~~ 1	 & ~~~~ 0.18$\,\pm\,$0.04	 & ~~~~ 71.0$\,\pm\,$36.0	 & ~~~~ 44.116$\,\pm\,$0.025 \ \\
 \ 021211	 & ~~~~ 1.01	 & ~~~~ 0.42$\,\pm\,$0.05	 & ~~~~ 127.0$\,\pm\,$52.0	 & ~~~~ 44.143$\,\pm\,$0.026 \ \\
 \ 991216	 & ~~~~ 1.02	 & ~~~~ 24.8$\,\pm\,$2.5	 & ~~~~ 648.0$\,\pm\,$134.0	 & ~~~~ 44.171$\,\pm\,$0.026 \ \\
 \ 080411	 & ~~~~ 1.03	 & ~~~~ 5.7$\,\pm\,$0.3	 & ~~~~ 524.0$\,\pm\,$70.0	 & ~~~~ 44.198$\,\pm\,$0.027 \ \\
 \ 000911	 & ~~~~ 1.06	 & ~~~~ 23.0$\,\pm\,$4.7	 & ~~~~ 1856.0$\,\pm\,$371.0	 & ~~~~ 44.279$\,\pm\,$0.029 \ \\
 \ 091208B	 & ~~~~ 1.063	 & ~~~~ 0.79$\,\pm\,$0.06	 & ~~~~ 255.0$\,\pm\,$25.0	 & ~~~~ 44.287$\,\pm\,$0.029 \ \\
 \ 091024	 & ~~~~ 1.092	 & ~~~~ 16.57$\,\pm\,$1.6	 & ~~~~ 586.0$\,\pm\,$251.0	 & ~~~~ 44.364$\,\pm\,$0.031 \ \\
 \ 980613	 & ~~~~ 1.096	 & ~~~~ 0.19$\,\pm\,$0.03	 & ~~~~ 194.0$\,\pm\,$89.0	 & ~~~~ 44.374$\,\pm\,$0.032 \ \\
 \ 080413B	 & ~~~~ 1.1	 & ~~~~ 0.73$\,\pm\,$0.09	 & ~~~~ 150.0$\,\pm\,$30.0	 & ~~~~ 44.385$\,\pm\,$0.032 \ \\
 \ 000418	 & ~~~~ 1.12	 & ~~~~ 2.8$\,\pm\,$0.5	 & ~~~~ 284.0$\,\pm\,$21.0	 & ~~~~ 44.436$\,\pm\,$0.034 \ \\
 \ 061126	 & ~~~~ 1.1588	 & ~~~~ 8.7$\,\pm\,$1.0	 & ~~~~ 1337.0$\,\pm\,$410.0	 & ~~~~ 44.535$\,\pm\,$0.037 \ \\
 \ 090926B	 & ~~~~ 1.24	 & ~~~~ 0.83$\,\pm\,$0.04	 & ~~~~ 204.0$\,\pm\,$10.0	 & ~~~~ 44.734$\,\pm\,$0.046 \ \\
 \ 020813	 & ~~~~ 1.25	 & ~~~~ 16.3$\,\pm\,$4.1	 & ~~~~ 590.0$\,\pm\,$151.0	 & ~~~~ 44.758$\,\pm\,$0.047 \ \\
 \ 061007	 & ~~~~ 1.261	 & ~~~~ 21.1$\,\pm\,$2.1	 & ~~~~ 890.0$\,\pm\,$124.0	 & ~~~~ 44.784$\,\pm\,$0.048 \ \\
 \ 050126A	 & ~~~~ 1.29	 & ~~~~ 0.17$\,\pm\,$0.04	 & ~~~~ 263.0$\,\pm\,$110.0	 & ~~~~ 44.852$\,\pm\,$0.052 \ \\
 \ 990506	 & ~~~~ 1.3	 & ~~~~ 21.7$\,\pm\,$2.2	 & ~~~~ 677.0$\,\pm\,$156.0	 & ~~~~ 44.876$\,\pm\,$0.053 \ \\
 \ 061121	 & ~~~~ 1.314	 & ~~~~ 5.1$\,\pm\,$0.6	 & ~~~~ 1289.0$\,\pm\,$153.0	 & ~~~~ 44.908$\,\pm\,$0.055 \ \\
 \ 071117	 & ~~~~ 1.331	 & ~~~~ 0.89$\,\pm\,$0.21	 & ~~~~ 647.0$\,\pm\,$226.0	 & ~~~~ 44.947$\,\pm\,$0.057 \ \\
 \ 100414A	 & ~~~~ 1.368	 & ~~~~ 15.99$\,\pm\,$0.25	 & ~~~~ 1486.16$\,\pm\,$29.6	 & ~~~~ 45.031$\,\pm\,$0.062 \ \\[1mm]
 \hline\hline
 \end{tabular}
 \end{center}
 \caption{\label{tab1} The numerical data of 59 low-redshift
 GRBs at $z<1.4$. The first 4 columns are taken from~\cite{r16,r23},
 whereas the last column is derived by using Pad\'e method.
 These 59 low-redshift GRBs can be used to calibrate the Amati
 relation. See the text for details.}
 \end{table}
 \endgroup


\vspace{-12mm} 

The distance moduli of the 59 low-redshift GRBs at $z_i<1.4$
 can be directly read from the formula $\mu_{pade}(z_i)$ in
 Eq.~(\ref{eq4}) with the best-fit coefficients.
 The corresponding errors can be obtained by using the
 well-known error propagation equation for any quantity
 $Q(x_i)$~\cite{r25} (see also e.g.~\cite{r26,r27})
 \be{eq6}
 \sigma^2(Q)=\sum\limits_i^n \left(
 \frac{\partial Q}{\partial x_i}\right)^2_{x=\bar{x}}C_{ii}
 +2\sum\limits_{i=1}^n\sum\limits_{j=i+1}^n
 \left(\frac{\partial Q}{\partial x_i}\frac{\partial Q}
 {\partial x_j}\right)_{x=\bar{x}}C_{ij}\,,
 \ee
 where $C$ is the covariance matrix. In our case, the corresponding
 covariance matrix is given in Eq.~(\ref{eq5}). We plot the derived
 distance moduli $\mu$ and the corresponding error bars of these 59
 low-redshift GRBs in the left panel of Fig.~\ref{fig1inst}. In
 Table~\ref{tab1}, we present the numerical data of these 59
 low-redshift GRBs.

As is well known, based on a sample of 12 BeppoSAX GRBs with
 known redshift, Amati {\it et al.}~\cite{r28} found an
 empirical relation between the cosmological rest-frame
 spectrum peak energy $E_{\rm p,i}=E_{\rm p,obs}\times(1+z)$
 and the isotropic equivalent radiated energy $E_{\rm iso}$,
 namely $E_{\rm p,i}=K\times E_{\rm iso}^m$. Note that the
 isotropic equivalent radiated energy is given by
 \be{eq7}
 E_{\rm iso}=4\pi\,d_L^2\,S_{\rm bolo}\,(1+z)^{-1},
 \ee
 where $S_{\rm bolo}$ is the bolometric fluence of gamma
 rays in the GRB at redshift $z$, and $d_L$ is the luminosity
 distance of the GRB. Following e.g.~\cite{r15,r16}, in this
 work we calibrate GRBs with the Amati relation.
 For convenience, following e.g.~\cite{r29,r16}, we recast
 the Amati relation as
 \be{eq8}
 \log\frac{E_{\rm iso}}{\rm erg}=\lambda+b\,
 \log\frac{E_{\rm p,i}}{\rm \,300\,keV\,}\,,
 \ee
 where ``$\log$'' indicates the logarithm to base $10$,
 whereas $\lambda$ and $b$ are constants to be determined.
 By using the well-known relation (see e.g.~\cite{r1,r59,r60}
 for several pedagogical textbooks. Note that $d_L$ is in units
 of Mpc, and $m$, $M$ are the apparent magnitude and the
 absolute magnitude, respectively)
 \be{eq9}
 \mu\equiv m-M=5\log\frac{d_L}{\rm \,Mpc}+25\,,
 \ee
 we can convert the distance modulus $\mu$ of each low-redshift
 GRB into luminosity distance $d_L$ (in units of Mpc), and then
 $E_{\rm iso}$ by employing Eq.~(\ref{eq7}) while $S_{\rm bolo}$ is
 known in~\cite{r16,r23}. We present them in the right panel of
 Fig.~\ref{fig1inst}, whereas $E_{\rm p,i}$ of these 59
 low-redshift GRBs at $z<1.4$ are taken from~\cite{r16,r23}.
 From Fig.~\ref{fig1inst}, one can clearly see that the intrinsic
 scatter is dominating over the measurement errors. Therefore,
 as in~\cite{r29,r14}, the bisector of the two ordinary least
 squares~\cite{r30} will be used. Following the procedure of
 the bisector of the two ordinary least squares described
 in~\cite{r30}, we find the best fit to be
 \be{eq10}
 b=1.7969~~~~~~~{\rm and}~~~~~~~\lambda=52.7333\,,
 \ee
 with $1\sigma$ uncertainties
 \be{eq11}
 \sigma_b=0.0070~~~~~~~{\rm and}~~~~~~~\sigma_\lambda=0.0035\,.
 \ee
 The best-fit calibration line Eq.~(\ref{eq8}) with $b$ and
 $\lambda$ in Eq.~(\ref{eq10}) is also plotted in the right
 panel of Fig.~\ref{fig1inst}. From Eq.~(\ref{eq11}), one
 can see that the calibration in this work is slightly better
 than the one in~\cite{r15,r16}.

\begin{center}
 \begin{figure}[b]
 \centering
 \includegraphics[width=0.5\textwidth]{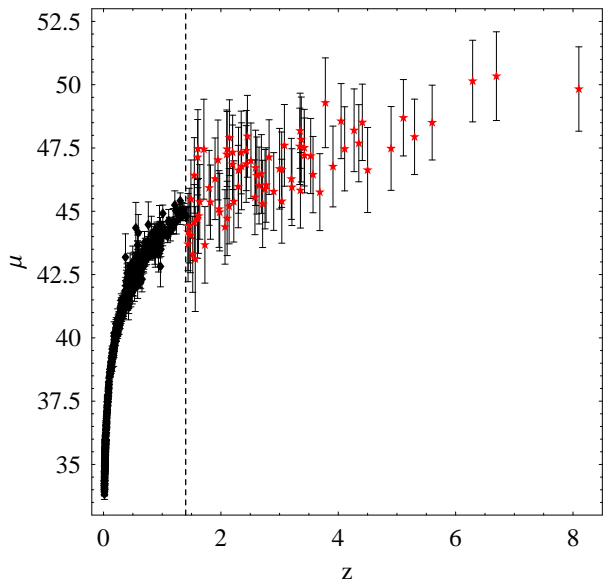}
 \caption{\label{fig2}
 The Hubble diagram of 580 Union2.1 SNIa (black diamonds)
 and 79 high-redshift GRBs (red stars) whose distance moduli
 are derived by using the calibrated Amati relation. The dashed
 line indicates $z=1.4$. See the text for details.}
 \end{figure}
 \end{center}

Next, we extend the calibrated Amati relation to
 high redshift, namely $z>1.4$. Since $E_{\rm p,i}$ for the
 79 GRBs at $z>1.4$ have been given in~\cite{r16,r23}, we can
 derive $E_{\rm iso}$ from the calibrated Amati relation
 Eq.~(\ref{eq8}) with $b$ and $\lambda$ in Eq.~(\ref{eq10}).
 Then, we derive the distance moduli $\mu$ for these 79 GRBs
 at $z>1.4$ using Eqs.~(\ref{eq7}) and (\ref{eq9}) while
 their $S_{\rm bolo}$ can be taken from~\cite{r16,r23}. On
 the other hand, the propagated uncertainties are
 given by~\cite{r29}
 \be{eq12}
 \sigma_\mu=\left[\left(\frac{5}{2}\sigma_{\log E_{\rm iso}}
 \right)^2+\left(\frac{5}{2\ln 10}\,
 \frac{\sigma_{S_{\rm bolo}}}{S_{\rm bolo}}
 \right)^2\right]^{1/2}\,,
 \ee
 where
 \be{eq13}
 \sigma_{\log E_{\rm iso}}^2=\sigma_\lambda^2+\left(\sigma_b
 \log\frac{E_{\rm p,i}}{\rm \,300\,keV\,}\right)^2+\left(
 \frac{b}{\ln 10}\,\frac{\sigma_{E_{\rm p,i}}}{E_{\rm p,i}}
 \right)^2
 +\sigma_{E_{\rm iso,sys}}^2\,,
 \ee
 in which $\sigma_{E_{\rm iso,sys}}$ is the systematic error
 and it accounts the extra scatter of the luminosity relation.
 As in~\cite{r29}, by requiring the $\chi^2/dof$ of the
 59 points at $z<1.4$ in the
 $\log E_{\rm p,i}\,/({\rm 300\,keV})-\log E_{\rm iso}\,/{\rm erg}$
 plane about the best-fit calibration line to be
 unity, we find that
 \be{eq14}
 \sigma_{E_{\rm iso,sys}}^2=0.1547\,.
 \ee
 Note that in principle $\sigma_{E_{\rm iso,sys}}^2$ is
 a free parameter. However, if we allow it
 to vary with cosmology, as in e.g.~\cite{r31},
 there might be a room for the circularity problem. Even
 if one does not care this problem, the constraints on cosmological
 models become loose, mainly due to the fact that
 the number of free parameters has been increased. On the
 other hand, we have not used any cosmology when
 we calibrate GRBs at $z<1.4$, so we have no freedom to
 determine $\sigma_{E_{\rm iso,sys}}^2$ by cosmology, and
 hence we should use the method in~\cite{r29} to
 fix it by requiring $\chi^2/dof=1$. We admit that this
 prevents us to learn the systematics dominating
 the Amati relation. Anyway, we plot the derived distance
 moduli $\mu$ with $1\sigma$ uncertainties for these 79
 GRBs at $z>1.4$ in Fig.~\ref{fig2}. We also present the
 numerical data of these 79 high-redshift GRBs in Table~\ref{tab2}.
 It is worth noting that these 79 high-redshift GRBs are
 obtained in a completely cosmology-independent manner,
 and hence can be used to constrain cosmological models without
 the circularity problem. We name them Mayflower sample
 for convenience.


 \begingroup
 \squeezetable
 \begin{table}[ptbh]
 \begin{center}
 \begin{tabular}{llccc}
 \hline \hline \ GRB & ~~~~~$z$ & ~~~~$S_{\rm bolo}~(10^{-5}~\rm erg\, cm^{-2})$
 & ~~~~~$E_{\rm p,i}~(\rm keV)$ & \ \ $\mu$ \\[1mm] \hline
 \ 100814A	 & ~~~~ 1.44	 & ~~~~ 2.8$\,\pm\,$0.09	 & ~~~~ 259.616$\,\pm\,$33.92	 & ~~~~ 43.71$\,\pm\,$1.49 \ \\[-1.2mm]
 \ 050318	 & ~~~~ 1.44	 & ~~~~ 0.42$\,\pm\,$0.03	 & ~~~~ 115.0$\,\pm\,$25.0	 & ~~~~ 44.18$\,\pm\,$1.53 \ \\[-1.2mm]
 \ 110213A	 & ~~~~ 1.46	 & ~~~~ 1.27$\,\pm\,$0.04	 & ~~~~ 242.064$\,\pm\,$20.91	 & ~~~~ 44.43$\,\pm\,$1.47 \ \\[-1.2mm]
 \ 010222	 & ~~~~ 1.48	 & ~~~~ 14.6$\,\pm\,$1.5	 & ~~~~ 766.0$\,\pm\,$30.0	 & ~~~~ 44.04$\,\pm\,$1.47 \ \\[-1.2mm]
 \ 060418	 & ~~~~ 1.489	 & ~~~~ 2.3$\,\pm\,$0.5	 & ~~~~ 572.0$\,\pm\,$143.0	 & ~~~~ 45.48$\,\pm\,$1.54 \ \\[-1.2mm]
 \ 030328	 & ~~~~ 1.52	 & ~~~~ 6.4$\,\pm\,$0.6	 & ~~~~ 328.0$\,\pm\,$55.0	 & ~~~~ 43.3$\,\pm\,$1.5 \ \\[-1.2mm]
 \ 070125	 & ~~~~ 1.547	 & ~~~~ 13.3$\,\pm\,$1.3	 & ~~~~ 934.0$\,\pm\,$148.0	 & ~~~~ 44.56$\,\pm\,$1.5 \ \\[-1.2mm]
 \ 090102	 & ~~~~ 1.547	 & ~~~~ 3.48$\,\pm\,$0.63	 & ~~~~ 1149.0$\,\pm\,$166.0	 & ~~~~ 46.42$\,\pm\,$1.49 \ \\[-1.2mm]
 \ 040912	 & ~~~~ 1.563	 & ~~~~ 0.21$\,\pm\,$0.06	 & ~~~~ 44.0$\,\pm\,$33.0	 & ~~~~ 43.11$\,\pm\,$2.07 \ \\[-1.2mm]
 \ 990123	 & ~~~~ 1.6	 & ~~~~ 35.8$\,\pm\,$5.8	 & ~~~~ 1724.0$\,\pm\,$466.0	 & ~~~~ 44.7$\,\pm\,$1.56 \ \\[-1.2mm]
 \ 071003	 & ~~~~ 1.604	 & ~~~~ 5.32$\,\pm\,$0.59	 & ~~~~ 2077.0$\,\pm\,$286.0	 & ~~~~ 47.14$\,\pm\,$1.49 \ \\[-1.2mm]
 \ 090418	 & ~~~~ 1.608	 & ~~~~ 2.35$\,\pm\,$0.59	 & ~~~~ 1567.0$\,\pm\,$384.0	 & ~~~~ 47.48$\,\pm\,$1.54 \ \\[-1.2mm]
 \ 990510	 & ~~~~ 1.619	 & ~~~~ 2.6$\,\pm\,$0.4	 & ~~~~ 423.0$\,\pm\,$42.0	 & ~~~~ 44.82$\,\pm\,$1.48 \ \\[-1.2mm]
 \ 080605	 & ~~~~ 1.6398	 & ~~~~ 3.4$\,\pm\,$0.28	 & ~~~~ 650.0$\,\pm\,$55.0	 & ~~~~ 45.37$\,\pm\,$1.47 \ \\[-1.2mm]
 \ 091020	 & ~~~~ 1.71	 & ~~~~ 0.11$\,\pm\,$0.03	 & ~~~~ 280.0$\,\pm\,$190.0	 & ~~~~ 47.45$\,\pm\,$1.97 \ \\[-1.2mm]
 \ 100906A	 & ~~~~ 1.727	 & ~~~~ 3.91$\,\pm\,$0.04	 & ~~~~ 289.062$\,\pm\,$55.0854	 & ~~~~ 43.67$\,\pm\,$1.51 \ \\[-1.2mm]
 \ 080514B	 & ~~~~ 1.8	 & ~~~~ 2.03$\,\pm\,$0.48	 & ~~~~ 627.0$\,\pm\,$65.0	 & ~~~~ 45.93$\,\pm\,$1.48 \ \\[-1.2mm]
 \ 090902B	 & ~~~~ 1.822	 & ~~~~ 32.38$\,\pm\,$1.01	 & ~~~~ 2187.0$\,\pm\,$31.0	 & ~~~~ 45.36$\,\pm\,$1.47 \ \\[-1.2mm]
 \ 020127	 & ~~~~ 1.9	 & ~~~~ 0.38$\,\pm\,$0.01	 & ~~~~ 290.0$\,\pm\,$100.0	 & ~~~~ 46.28$\,\pm\,$1.61 \ \\[-1.2mm]
 \ 080319C	 & ~~~~ 1.95	 & ~~~~ 1.5$\,\pm\,$0.3	 & ~~~~ 906.0$\,\pm\,$272.0	 & ~~~~ 47.03$\,\pm\,$1.58 \ \\[-1.2mm]
 \ 081008	 & ~~~~ 1.9685	 & ~~~~ 0.96$\,\pm\,$0.09	 & ~~~~ 261.0$\,\pm\,$52.0	 & ~~~~ 45.09$\,\pm\,$1.52 \ \\[-1.2mm]
 \ 030226	 & ~~~~ 1.98	 & ~~~~ 1.3$\,\pm\,$0.1	 & ~~~~ 289.0$\,\pm\,$66.0	 & ~~~~ 44.97$\,\pm\,$1.53 \ \\[-1.2mm]
 \ 000926	 & ~~~~ 2.07	 & ~~~~ 2.6$\,\pm\,$0.6	 & ~~~~ 310.0$\,\pm\,$20.0	 & ~~~~ 44.38$\,\pm\,$1.47 \ \\[-1.2mm]
 \ 081203A	 & ~~~~ 2.1	 & ~~~~ 3.33$\,\pm\,$0.29	 & ~~~~ 1541.0$\,\pm\,$757.0	 & ~~~~ 47.25$\,\pm\,$1.75 \ \\[-1.2mm]
 \ 100728B	 & ~~~~ 2.106	 & ~~~~ 0.25$\,\pm\,$0.01	 & ~~~~ 406.886$\,\pm\,$46.59	 & ~~~~ 47.46$\,\pm\,$1.48 \ \\[-1.2mm]
 \ 090926	 & ~~~~ 2.1062	 & ~~~~ 15.08$\,\pm\,$0.77	 & ~~~~ 974.0$\,\pm\,$50.0	 & ~~~~ 44.72$\,\pm\,$1.47 \ \\[-1.2mm]
 \ 011211	 & ~~~~ 2.14	 & ~~~~ 0.5$\,\pm\,$0.06	 & ~~~~ 186.0$\,\pm\,$24.0	 & ~~~~ 45.2$\,\pm\,$1.49 \ \\[-1.2mm]
 \ 071020	 & ~~~~ 2.145	 & ~~~~ 0.87$\,\pm\,$0.4	 & ~~~~ 1013.0$\,\pm\,$160.0	 & ~~~~ 47.91$\,\pm\,$1.5 \ \\[-1.2mm]
 \ 050922C	 & ~~~~ 2.198	 & ~~~~ 0.47$\,\pm\,$0.16	 & ~~~~ 415.0$\,\pm\,$111.0	 & ~~~~ 46.85$\,\pm\,$1.55 \ \\[-1.2mm]
 \ 080804	 & ~~~~ 2.2	 & ~~~~ 1.01$\,\pm\,$0.18	 & ~~~~ 810.0$\,\pm\,$45.0	 & ~~~~ 47.33$\,\pm\,$1.47 \ \\[-1.2mm]
 \ 110205A	 & ~~~~ 2.22	 & ~~~~ 4.84$\,\pm\,$0.52	 & ~~~~ 715.0$\,\pm\,$239.0	 & ~~~~ 45.39$\,\pm\,$1.6 \ \\[-1.2mm]
 \ 060124	 & ~~~~ 2.296	 & ~~~~ 3.4$\,\pm\,$0.5	 & ~~~~ 784.0$\,\pm\,$285.0	 & ~~~~ 45.98$\,\pm\,$1.63 \ \\[-1.2mm]
 \ 021004	 & ~~~~ 2.3	 & ~~~~ 0.27$\,\pm\,$0.04	 & ~~~~ 266.0$\,\pm\,$117.0	 & ~~~~ 46.62$\,\pm\,$1.7 \ \\[-1.2mm]
 \ 051109A	 & ~~~~ 2.346	 & ~~~~ 0.51$\,\pm\,$0.05	 & ~~~~ 539.0$\,\pm\,$200.0	 & ~~~~ 47.32$\,\pm\,$1.63 \ \\[-1.2mm]
 \ 070110	 & ~~~~ 2.352	 & ~~~~ 0.43$\,\pm\,$0.12	 & ~~~~ 370.0$\,\pm\,$170.0	 & ~~~~ 46.78$\,\pm\,$1.72 \ \\[-1.2mm]
 \ 060908	 & ~~~~ 2.43	 & ~~~~ 0.73$\,\pm\,$0.07	 & ~~~~ 514.0$\,\pm\,$102.0	 & ~~~~ 46.87$\,\pm\,$1.52 \ \\[-1.2mm]
 \ 080413	 & ~~~~ 2.433	 & ~~~~ 0.56$\,\pm\,$0.14	 & ~~~~ 584.0$\,\pm\,$180.0	 & ~~~~ 47.4$\,\pm\,$1.58 \ \\[-1.2mm]
 \ 090812	 & ~~~~ 2.452	 & ~~~~ 3.08$\,\pm\,$0.53	 & ~~~~ 2000.0$\,\pm\,$700.0	 & ~~~~ 47.96$\,\pm\,$1.62 \ \\[-1.2mm]
 \ 081121	 & ~~~~ 2.512	 & ~~~~ 1.71$\,\pm\,$0.33	 & ~~~~ 871.0$\,\pm\,$123.0	 & ~~~~ 47.0$\,\pm\,$1.49 \ \\[-1.2mm]
 \ 081118	 & ~~~~ 2.58	 & ~~~~ 0.27$\,\pm\,$0.06	 & ~~~~ 147.0$\,\pm\,$14.0	 & ~~~~ 45.55$\,\pm\,$1.48 \ \\[-1.2mm]
 \ 080721	 & ~~~~ 2.591	 & ~~~~ 7.86$\,\pm\,$1.37	 & ~~~~ 1741.0$\,\pm\,$227.0	 & ~~~~ 46.72$\,\pm\,$1.49 \ \\[-1.2mm]
 \ 050820	 & ~~~~ 2.612	 & ~~~~ 6.4$\,\pm\,$0.5	 & ~~~~ 1325.0$\,\pm\,$277.0	 & ~~~~ 46.42$\,\pm\,$1.52 \ \\[-1.2mm]
 \ 030429	 & ~~~~ 2.65	 & ~~~~ 0.14$\,\pm\,$0.02	 & ~~~~ 128.0$\,\pm\,$26.0	 & ~~~~ 46.02$\,\pm\,$1.52 \ \\[-1.2mm]
 \ 080603B	 & ~~~~ 2.69	 & ~~~~ 0.64$\,\pm\,$0.06	 & ~~~~ 376.0$\,\pm\,$100.0	 & ~~~~ 46.48$\,\pm\,$1.55 \ \\[-1.2mm]
 \ 060714	 & ~~~~ 2.711	 & ~~~~ 0.82$\,\pm\,$0.06	 & ~~~~ 234.0$\,\pm\,$109.0	 & ~~~~ 45.29$\,\pm\,$1.72 \ \\[-1.2mm]
 \ 091029	 & ~~~~ 2.752	 & ~~~~ 0.47$\,\pm\,$0.04	 & ~~~~ 230.0$\,\pm\,$66.0	 & ~~~~ 45.87$\,\pm\,$1.57 \ \\[-1.2mm]
 \ 081222	 & ~~~~ 2.77	 & ~~~~ 1.67$\,\pm\,$0.17	 & ~~~~ 505.0$\,\pm\,$34.0	 & ~~~~ 46.04$\,\pm\,$1.47 \ \\[-1.2mm]
 \ 050603	 & ~~~~ 2.821	 & ~~~~ 3.5$\,\pm\,$0.2	 & ~~~~ 1333.0$\,\pm\,$107.0	 & ~~~~ 47.14$\,\pm\,$1.47 \ \\[-1.2mm]
 \ 050401	 & ~~~~ 2.9	 & ~~~~ 1.9$\,\pm\,$0.4	 & ~~~~ 467.0$\,\pm\,$110.0	 & ~~~~ 45.78$\,\pm\,$1.54 \ \\[-1.2mm]
 \ 090715B	 & ~~~~ 3	 & ~~~~ 1.09$\,\pm\,$0.17	 & ~~~~ 536.0$\,\pm\,$172.0	 & ~~~~ 46.68$\,\pm\,$1.59 \ \\[-1.2mm]
 \ 080607	 & ~~~~ 3.036	 & ~~~~ 8.96$\,\pm\,$0.48	 & ~~~~ 1691.0$\,\pm\,$226.0	 & ~~~~ 46.65$\,\pm\,$1.49 \ \\[-1.2mm]
 \ 081028	 & ~~~~ 3.038	 & ~~~~ 0.81$\,\pm\,$0.1	 & ~~~~ 234.0$\,\pm\,$93.0	 & ~~~~ 45.4$\,\pm\,$1.66 \ \\[-1.2mm]
 \ 060607A	 & ~~~~ 3.082	 & ~~~~ 0.54$\,\pm\,$0.08	 & ~~~~ 575.0$\,\pm\,$200.0	 & ~~~~ 47.61$\,\pm\,$1.61 \ \\[-1.2mm]
 \ 020124	 & ~~~~ 3.2	 & ~~~~ 1.2$\,\pm\,$0.1	 & ~~~~ 448.0$\,\pm\,$148.0	 & ~~~~ 46.28$\,\pm\,$1.6 \ \\[-1.2mm]
 \ 060526	 & ~~~~ 3.21	 & ~~~~ 0.12$\,\pm\,$0.06	 & ~~~~ 105.0$\,\pm\,$21.0	 & ~~~~ 45.95$\,\pm\,$1.52 \ \\[-1.2mm]
 \ 050908	 & ~~~~ 3.344	 & ~~~~ 0.09$\,\pm\,$0.01	 & ~~~~ 195.0$\,\pm\,$36.0	 & ~~~~ 47.57$\,\pm\,$1.51 \ \\[-1.2mm]
 \ 080810	 & ~~~~ 3.35	 & ~~~~ 1.82$\,\pm\,$0.2	 & ~~~~ 1470.0$\,\pm\,$180.0	 & ~~~~ 48.18$\,\pm\,$1.48 \ \\[-1.2mm]
 \ 061222B	 & ~~~~ 3.355	 & ~~~~ 0.44$\,\pm\,$0.07	 & ~~~~ 200.0$\,\pm\,$28.0	 & ~~~~ 45.83$\,\pm\,$1.49 \ \\[-1.2mm]
 \ 030323	 & ~~~~ 3.37	 & ~~~~ 0.12$\,\pm\,$0.04	 & ~~~~ 270.0$\,\pm\,$113.0	 & ~~~~ 47.84$\,\pm\,$1.68 \ \\[-1.2mm]
 \ 971214	 & ~~~~ 3.42	 & ~~~~ 0.87$\,\pm\,$0.11	 & ~~~~ 685.0$\,\pm\,$133.0	 & ~~~~ 47.51$\,\pm\,$1.51 \ \\[-1.2mm]
 \ 060707	 & ~~~~ 3.425	 & ~~~~ 0.23$\,\pm\,$0.04	 & ~~~~ 279.0$\,\pm\,$28.0	 & ~~~~ 47.21$\,\pm\,$1.48 \ \\[-1.2mm]
 \ 060115	 & ~~~~ 3.53	 & ~~~~ 0.25$\,\pm\,$0.04	 & ~~~~ 285.0$\,\pm\,$34.0	 & ~~~~ 47.18$\,\pm\,$1.48 \ \\[-1.2mm]
 \ 090323	 & ~~~~ 3.57	 & ~~~~ 14.98$\,\pm\,$1.83	 & ~~~~ 1901.0$\,\pm\,$343.0	 & ~~~~ 46.45$\,\pm\,$1.51 \ \\[-1.2mm]
 \ 060906	 & ~~~~ 3.686	 & ~~~~ 0.55$\,\pm\,$0.06	 & ~~~~ 209.0$\,\pm\,$43.0	 & ~~~~ 45.76$\,\pm\,$1.52 \ \\[-1.2mm]
 \ 060605	 & ~~~~ 3.78	 & ~~~~ 0.1$\,\pm\,$0.02	 & ~~~~ 490.0$\,\pm\,$251.0	 & ~~~~ 49.29$\,\pm\,$1.77 \ \\[-1.2mm]
 \ 060210	 & ~~~~ 3.91	 & ~~~~ 1.4$\,\pm\,$0.19	 & ~~~~ 575.0$\,\pm\,$186.0	 & ~~~~ 46.77$\,\pm\,$1.59 \ \\[-1.2mm]
 \ 060206	 & ~~~~ 4.048	 & ~~~~ 0.14$\,\pm\,$0.03	 & ~~~~ 394.0$\,\pm\,$46.0	 & ~~~~ 48.56$\,\pm\,$1.48 \ \\[-1.2mm]
 \ 090516	 & ~~~~ 4.109	 & ~~~~ 1.96$\,\pm\,$0.38	 & ~~~~ 971.0$\,\pm\,$390.0	 & ~~~~ 47.47$\,\pm\,$1.66 \ \\[-1.2mm]
 \ 050505	 & ~~~~ 4.27	 & ~~~~ 0.52$\,\pm\,$0.08	 & ~~~~ 661.0$\,\pm\,$245.0	 & ~~~~ 48.2$\,\pm\,$1.63 \ \\[-1.2mm]
 \ 080916C	 & ~~~~ 4.35	 & ~~~~ 10.13$\,\pm\,$2.13	 & ~~~~ 2646.0$\,\pm\,$566.0	 & ~~~~ 47.69$\,\pm\,$1.52 \ \\[-1.2mm]
 \ 060223A	 & ~~~~ 4.41	 & ~~~~ 0.12$\,\pm\,$0.02	 & ~~~~ 339.0$\,\pm\,$63.0	 & ~~~~ 48.51$\,\pm\,$1.51 \ \\[-1.2mm]
 \ 000131	 & ~~~~ 4.5	 & ~~~~ 4.7$\,\pm\,$0.8	 & ~~~~ 987.0$\,\pm\,$416.0	 & ~~~~ 46.63$\,\pm\,$1.68 \ \\[-1.2mm]
 \ 060510B	 & ~~~~ 4.9	 & ~~~~ 0.87$\,\pm\,$0.07	 & ~~~~ 575.0$\,\pm\,$227.0	 & ~~~~ 47.48$\,\pm\,$1.65 \ \\[-1.2mm]
 \ 060522	 & ~~~~ 5.11	 & ~~~~ 0.17$\,\pm\,$0.03	 & ~~~~ 427.0$\,\pm\,$79.0	 & ~~~~ 48.7$\,\pm\,$1.51 \ \\[-1.2mm]
 \ 050814	 & ~~~~ 5.3	 & ~~~~ 0.24$\,\pm\,$0.05	 & ~~~~ 339.0$\,\pm\,$47.0	 & ~~~~ 47.94$\,\pm\,$1.49 \ \\[-1.2mm]
 \ 060927	 & ~~~~ 5.6	 & ~~~~ 0.27$\,\pm\,$0.04	 & ~~~~ 475.0$\,\pm\,$47.0	 & ~~~~ 48.51$\,\pm\,$1.48 \ \\[-1.2mm]
 \ 050904	 & ~~~~ 6.29	 & ~~~~ 2.0$\,\pm\,$0.2	 & ~~~~ 3178.0$\,\pm\,$1094.0	 & ~~~~ 50.15$\,\pm\,$1.61 \ \\[-1.2mm]
 \ 080913	 & ~~~~ 6.695	 & ~~~~ 0.12$\,\pm\,$0.03	 & ~~~~ 710.0$\,\pm\,$350.0	 & ~~~~ 50.34$\,\pm\,$1.75 \ \\[-1.2mm]
 \ 090423	 & ~~~~ 8.1	 & ~~~~ 0.12$\,\pm\,$0.03	 & ~~~~ 491.0$\,\pm\,$200.0	 & ~~~~ 49.83$\,\pm\,$1.67 \ \\[0.5mm]
 \hline\hline
 \end{tabular}
 \end{center}
 \vspace{-2.45mm}
 \caption{\label{tab2} The numerical data of 79 calibrated GRBs
 at $z>1.4$. The first 4 columns are taken from~\cite{r16,r23},
 whereas the last column is derived by using the calibrated
 Amati relation. These 79 calibrated GRBs are named Mayflower
 sample, and can be used to constrain cosmological models
 without the circularity problem.}
 \end{table}
 \endgroup



 \begin{center}
 \begin{figure}[tbhp]
 \centering
 \includegraphics[width=1.0\textwidth]{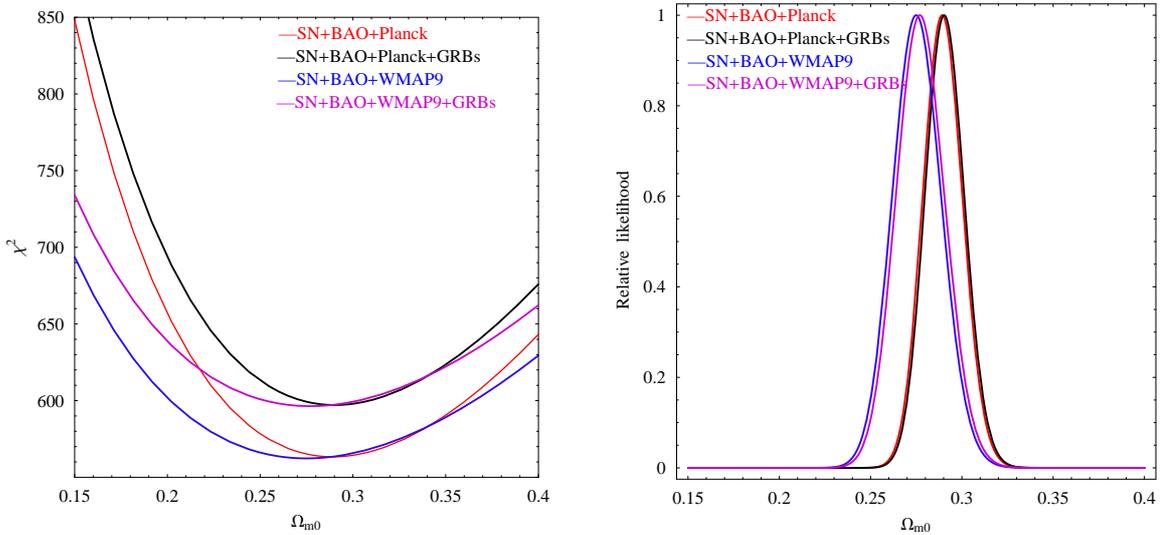}
 \caption{\label{fig3}
 The $\chi^2$ and likelihood ${\cal L}\propto e^{-\chi^2/2}$ as
 functions of $\Omega_{m0}$ from various joint datasets for the
 $\Lambda$CDM model.}
 \end{figure}
 \end{center}


\vspace{-12mm} 


\section{Observational constraints on cosmological models}\label{sec3}

In this section, we consider the observational constraints on
 various cosmological models. In addition to the 79 Mayflower
 GRBs obtained in the present work (see Table~\ref{tab2}), we
 also consider other types of observational data. Of course,
 the Union2.1 SNIa dataset~\cite{r24} (which consists of 580
 SNIa) will be used. The data ponits of both SNIa and GRBs
 are given in terms of the distance modulus. Following the
 methodology described in e.g.~\cite{r16} (especially, see the
 discussions between Eqs.~(9)---(13) of~\cite{r16} and note
 that $H_0$ has been marginalized), one can obtain the
 corresponding $\chi^2_{\rm SN}$ and $\chi^2_{\rm GRBs}$,
 respectively. Another important probe is the cosmic microwave
 background (CMB) anisotropy~\cite{r32,r33}. Recently, the WMAP
 Collaboration released their final 9-year data (WMAP9)~\cite{r32}.
 The Planck Collaboration also released their
 data (Planck)~\cite{r33}. As is noted by Planck Collaboration
 itself, there is a subtle tension between Planck data and
 WMAP data. So, in this work we consider these two CMB data
 separately. However, using the full data of CMB to perform a
 global fitting consumes a large amount of computation time
 and power. As an alternative, one can instead use the shift
 parameter $R$ from CMB data. It is argued in
 e.g.~\cite{r34,r35,r36} that it is model-independent and
 contains the main information of the full CMB data. The shift
 parameter $R$ is defined by~\cite{r34,r35,r36}
 \be{eq15}
 R\equiv\Omega_{m0}^{1/2}\int_0^{z_\ast}
 \frac{d\tilde{z}}{E(\tilde{z})}\,,
 \ee
 where $\Omega_{m0}$ is the present fractional density
 of pressureless matter; $z_\ast$ is the redshift
 of recombination; $E\equiv H/H_0$ and $H$ is the Hubble
 parameter (the subscript ``0'' indicates the present value
 of corresponding quantity). Its was determined in~\cite{r34}
 that $R=1.7302\pm 0.0169$ and $z_\ast=1089.09$
 for WMAP9~\cite{r32}, whereas $R=1.7499\pm 0.0088$ and
 $z_\ast=1090.41$ for Planck~\cite{r33}. The corresponding
 $\chi^2_{\rm CMB}=(R-R_{obs})^2/\sigma_R^2$. Finally, we
 consider also the observation of large-scale structure
 (LSS)~\cite{r37}. Similarly, it is also argued
 in e.g.~\cite{r35} that the distance parameter $A$ from the
 baryon acoustic oscillation (BAO) peak is model-independent
 and contains the main information of the LSS data. The
 distance parameter $A$ is defined by~\cite{r37,r37inst}
 \be{eq16}
 A\equiv\Omega_{m0}^{1/2}E(z_b)^{-1/3}\left[\frac{1}{z_b}
 \int_0^{z_b}\frac{d\tilde{z}}{E(\tilde{z})}\right]^{2/3},
 \ee
 where $z_b=0.35$. In~\cite{r37inst}, the value of $A$ has been
 determined to be $0.469\,(n_s/0.98)^{-0.35}\pm 0.017$. Here
 the scalar spectral index $n_s$ is taken to be
 $0.9662$~\cite{r36} from the Planck data~\cite{r33}. The
 corresponding $\chi^2_{\rm BAO}=(A-A_{obs})^2/\sigma_A^2$.
 Thus, the total $\chi^2$ is given by $\chi^2=\chi^2_{\rm SN}+
 \chi^2_{\rm BAO}+\chi^2_{\rm CMB}+\chi^2_{\rm GRBs}$.
 The best-fit model parameters are determined by
 minimizing the total $\chi^2$. As in~\cite{r39,r40}, the
 $68.3\%$ confidence level is determined by
 $\Delta\chi^2\equiv\chi^2-\chi^2_{min}\leq 1.0$, $2.3$,
 $3.53$, $4.72$ for $n_p=1$, $2$, $3$, $4$ respectively,
 where $n_p$ is the number of free model parameters. Similarly,
 the $95.4\%$ confidence level is determined by
 $\Delta\chi^2\equiv\chi^2-\chi^2_{min}\leq 4.0$, $6.18$,
 $8.02$, $9.72$ for $n_p=1$, $2$, $3$, $4$, respectively.

In the following subsections, we use various datasets to
 constrain cosmological models. To see the possible difference
 between Planck and WMAP9, we constrain the models with
 these two CMB data separately. We also consider the datasets
 with or without 79 Mayflower GRBs, to see the effect of GRBs
 on the constraints. So, in the followings, we use four
 joint datasets, namely, SN+BAO+Planck, SN+BAO+Planck+GRBs,
 SN+BAO+WMAP9, SN+BAO+WMAP9+GRBs, respectively. Note that
 we consider a flat Friedmann-Robertson-Walker (FRW) universe
 containing only pressureless matter and dark energy,
 except the case of Dvali-Gabadadze-Porrati (DGP) model in
 which the cosmic acceleration is due to a modification to
 general relativity rather than dark energy.


 \begin{table}[tb]
 \renewcommand{\arraystretch}{1.5}
 \begin{center}
 \begin{tabular}{l|cc|ccc} \hline\hline
 & \multicolumn{2}{c|}{$\Lambda$CDM Model}  &  \multicolumn{3}{c}{XCDM Model} \\
 \cline{2-6}    &     $\chi^2_{min}/dof$    &     $\Omega_{m0}$     &     $\chi^2_{min}/dof$
       &     $\Omega_{m0}$  & $~~~w_{\rm x}$        \\ \hline
  SN+BAO+Planck\ \ & 563.352/581 & \ $0.2892_{-0.0106}^{+0.0109}\,(1\sigma)\,_{-0.0208}^{+0.0222}\,(2\sigma)$
   \ & \ 562.645/580 \ &\ 0.2874 \ & \ $-1.0326$ \\ \hline
  SN+BAO+Planck+GRBs\ \ & \ 597.135/660 \ & \ $0.2903_{-0.0106}^{+0.0109}\,(1\sigma)\,_{-0.0208}^{+0.0221}\,(2\sigma)$
    \ & \ 596.545/659 \  & \ 0.2887 \ & \ $-1.0300$ \\ \hline
  SN+BAO+WMAP9\ \ & \ 562.325/581 \  & \ $0.2750_{-0.0131}^{+0.0135}\,(1\sigma)\,_{-0.0257}^{+0.0275}\,(2\sigma)$
   \ & \ 562.278/580 \  & \ 0.2742 \ & \ $-0.9900$ \\ \hline
  SN+BAO+WMAP9+GRBs\ \ & \ 596.396/660 \  & \ $0.2770_{-0.0130}^{+0.0135}\,(1\sigma)\,_{-0.0256}^{+0.0274}\,(2\sigma)$
   \ & \ 596.333/659 \  & \ 0.2761 \ & \ $-0.9883$ \\
 \hline\hline
 \end{tabular}
 \end{center}
 \caption{\label{tab3} The $\chi_{min}^2$ and the best-fit
 model parameters from various joint datasets for the $\Lambda$CDM,
 XCDM models, respectively. Note that the $1\sigma$ and
 $2\sigma$ uncertainties are also given for the $\Lambda$CDM
 model.}
 \end{table}



\subsection{$\Lambda$CDM model}\label{sec3a}

At first, we consider the observational constraints on the flat
 $\Lambda$CDM model. As is well known, the corresponding
 $E=H/H_0$ reads
 \be{eq17}
 E(z)=\sqrt{\Omega_{m0}(1+z)^3+(1-\Omega_{m0})}\,.
 \ee
 It is easy to obtain the total $\chi^2$ as a function of the
 single model parameter $\Omega_{m0}$ for the $\Lambda$CDM
 model. We present the corresponding $\chi^2$ and likelihood
 ${\cal L}\propto e^{-\chi^2/2}$ in Fig.~\ref{fig3}. In
 Table~\ref{tab3}, the $\chi_{min}^2$ and the best-fit model
 parameter $\Omega_{m0}$ (with $1\sigma$ and $2\sigma$
 uncertainties) from various joint datasets are given.
 From Fig.~\ref{fig3} and Table~\ref{tab3}, we can see that
 Planck data favors a larger $\Omega_{m0}$ than WMAP9 data,
 while GRBs data also favors a slightly larger $\Omega_{m0}$.


\subsection{XCDM model}\label{sec3b}

\begin{center}
 \begin{figure}[tbhp]
 \centering
 \includegraphics[width=1.0\textwidth]{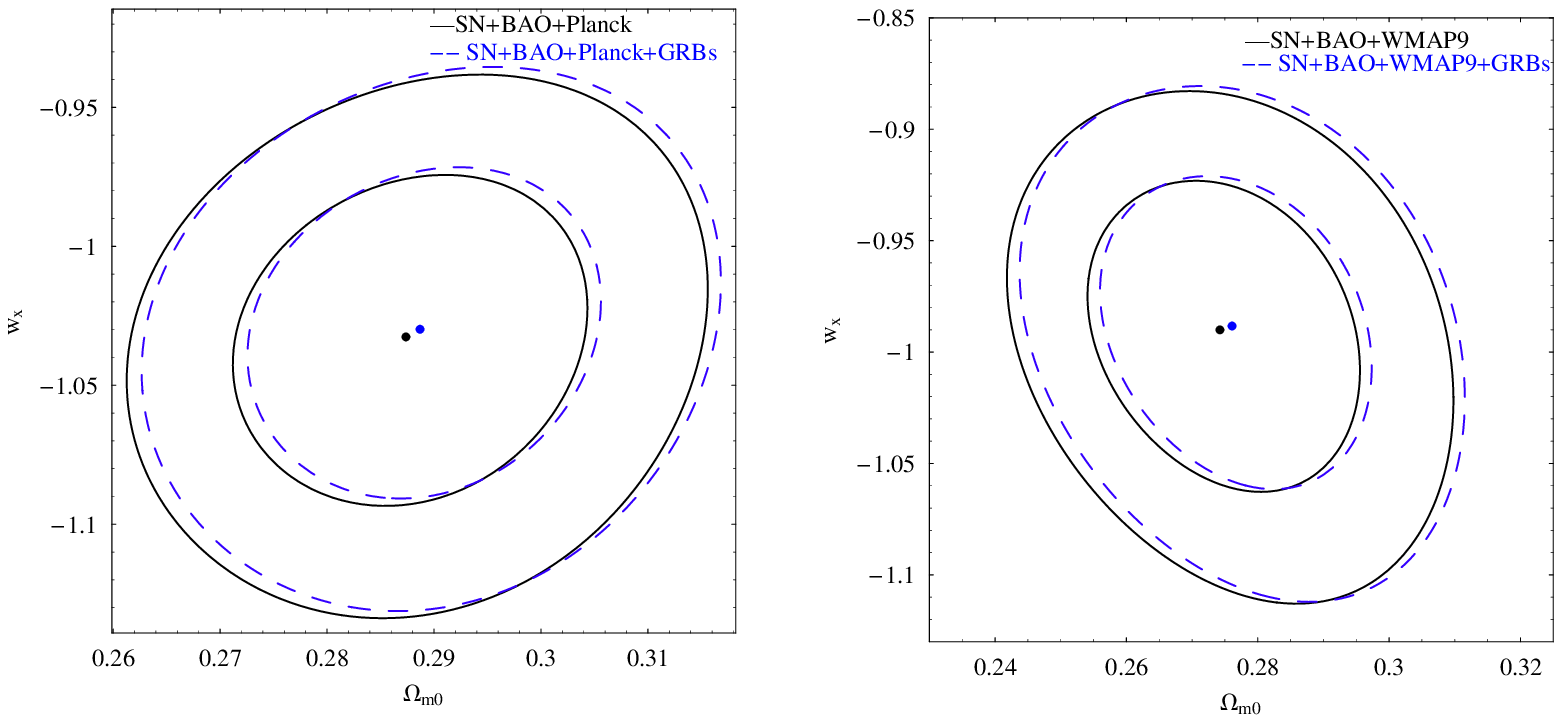}
 \caption{\label{fig4}
 The $68.3\%$ and $95.4\%$ confidence level contours in
 the $\Omega_{m0}-w_{\rm x}$ parameter space
 from various joint datasets for the XCDM model. The best-fit
 parameters are indicated by a solid point.}
 \end{figure}
 \end{center}

In the XCDM model, the equation-of-state parameter (EoS) of
 dark energy is a constant $w_{\rm x}$. The corresponding
 $E(z)$ is given by
 \be{eq18}
 E(z)=\sqrt{\Omega_{m0}(1+z)^3+
 (1-\Omega_{m0})(1+z)^{3(1+w_{\rm x})}}\,.
 \ee
 There are two free model parameters, namely $\Omega_{m0}$ and
 $w_{\rm x}$. By minimizing the corresponding total $\chi^2$,
 we find the best-fit parameters and present them
 in Table~\ref{tab3}. In Fig.~\ref{fig4}, we present the
 corresponding $68.3\%$ and $95.4\%$ confidence level contours
 in the $\Omega_{m0}-w_{\rm x}$ parameter space from various
 joint datasets for the XCDM model. From Fig.~\ref{fig4} and
 Table~\ref{tab3}, it is easy to see that Planck data favors a
 larger $\Omega_{m0}$ and a smaller $w_{\rm x}$ than WMAP9
 data, while GRBs data favors slightly larger values of both
 $\Omega_{m0}$ and $w_{\rm x}$.


 \begin{table}[tbhp]
 \renewcommand{\arraystretch}{1.5}
 \begin{center}
 \vspace{-7mm} 
 \begin{tabular}{l|ccccc} \hline\hline
  &  $\chi^2_{min}$  & $\chi^2_{min}/dof$ & $\ \Omega_{m0}$   & $~\ w_0$  &  $~~~w_a$ \\ \hline
  SN+BAO+Planck\ \ & \ \ 562.445 \ \  & 0.971 & \ \ \ 0.2847 \ \ \ &\ \ $-0.9775$\ \ & \ \ $-0.2782$ \ \\ \hline
  SN+BAO+Planck+GRBs\ \ & \ \ 596.480 \ \  & 0.907 & \ \ 0.2872 \ \ & \ \ $-0.9988$\ \ & \ \ $-0.1551$ \ \\ \hline
  SN+BAO+WMAP9\ \ & \ \ 562.235 \ \  & 0.971 & \ \ 0.2750 \ \ & \ \ \ $-1.0143$ \ \ & ~\ \ \ 0.1301 \ \\ \hline
  SN+BAO+WMAP9+GRBs~\ \ & \ \ 596.159 \ \  & 0.906 & \ \ 0.2776 \ \ & \ \ \ $-1.0363$ \ \ & ~\ \ \ 0.2546 \ \\
 \hline\hline
 \end{tabular}
 \end{center}
 \caption{\label{tab4} The $\chi_{min}^2$ and the best-fit
 model parameters from various joint datasets for the CPL
 model.}
 \end{table}


\vspace{-4mm} 


\subsection{CPL model}\label{sec3c}

In the well-known Chevallier-Polarski-Linder (CPL) model~\cite{r38},
 the EoS of dark energy is given by
 \be{eq19}
 w_{de}=w_0+w_a(1-a)=w_0+w_a\frac{z}{1+z}\,,
 \ee
 where $a$ is scale factor; $w_0$ and $w_a$ are
 both constants. The corresponding $E(z)$ is given by~\cite{r39,r40}
 \be{eq20}
 E(z)=\left[\Omega_{m0}(1+z)^3+
 \left(1-\Omega_{m0}\right)(1+z)^{3(1+w_0+w_a)}\exp\left(
 -\frac{3w_a z}{1+z}\right)\right]^{1/2}\,.
 \ee
 There are three free parameters in this model, namely
 $\Omega_{m0}$, $w_0$ and $w_a$. By minimizing the
 corresponding total $\chi^2$, we find the best-fit model
 parameters from various joint datasets for the CPL model, and
 present them in Table~\ref{tab4}. In Figs.~\ref{fig5},
 \ref{fig6} and \ref{fig7}, we also show the $68.3\%$
 and $95.4\%$ confidence level contours in the $w_0-w_a$,
 $\Omega_{m0}-w_0$ and $\Omega_{m0}-w_a$ planes, respectively.
 From Table~\ref{tab4} and Figs.~\ref{fig5}--\ref{fig7}, we
 see that Planck data favors a larger $\Omega_{m0}$, a larger
 $w_0$ and a smaller $w_a$ than WMAP9 data. Note that Planck
 data favors a $w_0>-1$ and a negative $w_a$, which means that
 dark energy was phantom-like ($w_{de}<-1$) in the past, then
 its EoS crossed the phantom divide, and became
 quintessence-like ($w_{de}>-1$) recently; finally its EoS will
 become positive in the future. On the contrary, WMAP9 data
 favors a $w_0<-1$ and a positive $w_a$, which means that
 dark energy was quintessence-like ($w_{de}>-1$) in the
 past, then its EoS crossed the phantom divide, and became
 phantom-like ($w_{de}<-1$) recently; finally the universe will
 end in a big rip. On the other hand, we find that GRBs data
 favors a slightly larger $\Omega_{m0}$, a slightly smaller
 $w_0$ and a slightly larger $w_a$.


 \begin{center}
 \vspace{3mm} 
 \begin{figure}[tbhp]
 \centering
 \includegraphics[width=1.0\textwidth]{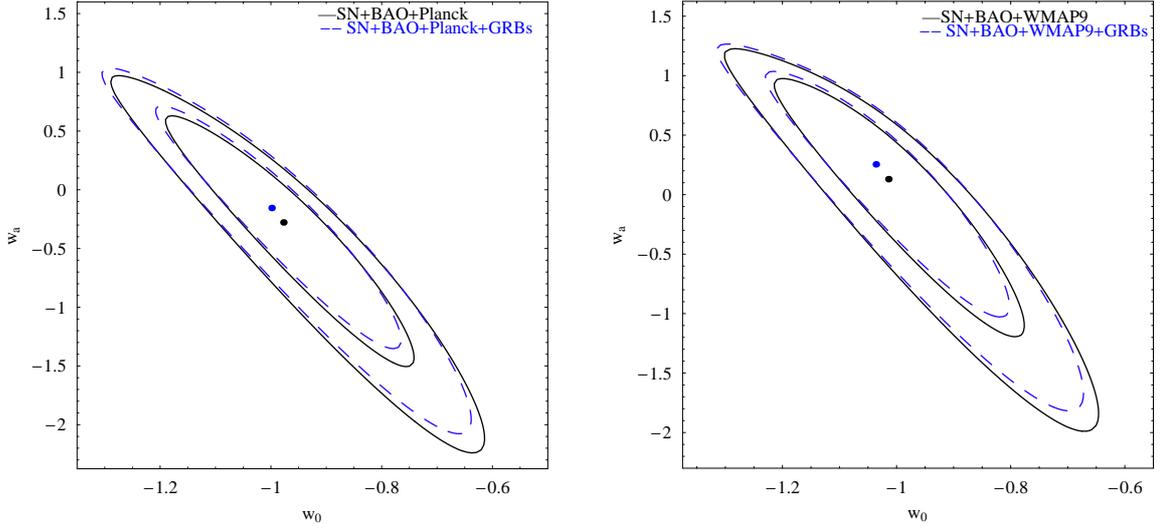}
 \caption{\label{fig5}
 The $68.3\%$ and $95.4\%$ confidence level contours in
 the $w_0-w_a$ plane from various joint datasets for the CPL model.
 The best-fit parameters are indicated by a solid point.}
 \end{figure}
 \end{center}



 \begin{center}
 \begin{figure}[tbhp]
 \centering
 \includegraphics[width=1.0\textwidth]{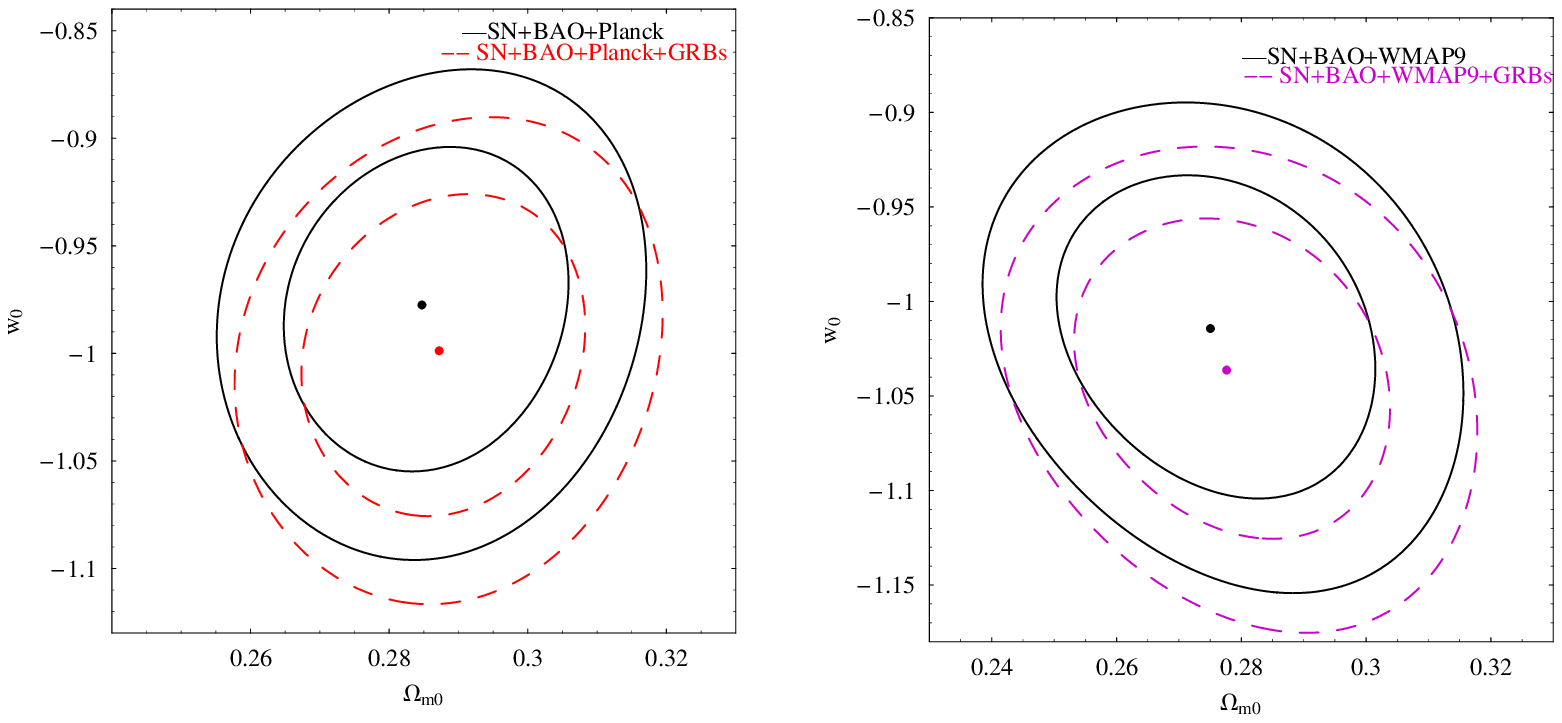}
 \caption{\label{fig6}
 The same as in Fig.~\ref{fig5}, except for the $\Omega_{m0}-w_0$ plane.}
 \end{figure}
 \end{center}



 \begin{center}
 \vspace{-12mm} 
 \begin{figure}[tbhp]
 \centering
 \includegraphics[width=1.0\textwidth]{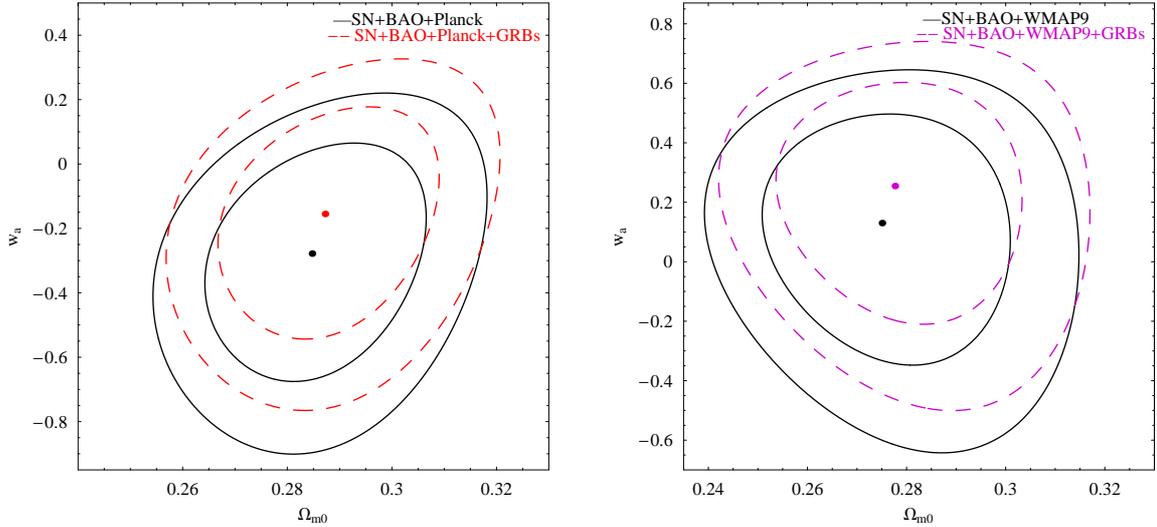}
 \caption{\label{fig7}
 The same as in Fig.~\ref{fig5}, except for the $\Omega_{m0}-w_a$ plane.}
 \end{figure}
 \end{center}


\vspace{-7mm} 


\subsection{DGP model}\label{sec3d}

\begin{center}
 \begin{figure}[tbhp]
 \centering
 \includegraphics[width=1.0\textwidth]{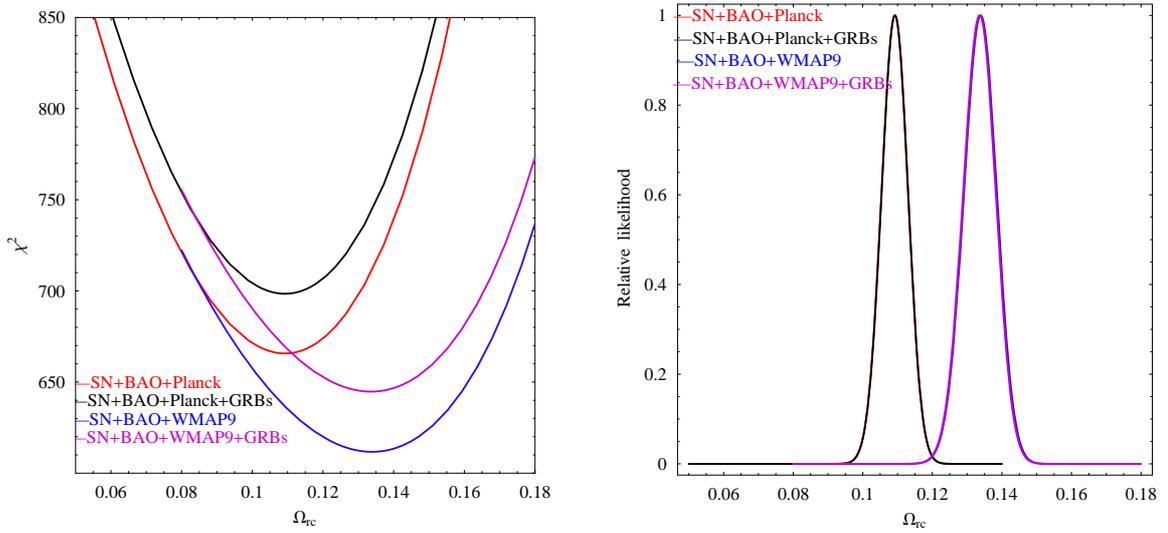}
 \caption{\label{fig8}
 The $\chi^2$ and likelihood ${\cal L}\propto e^{-\chi^2/2}$ as
 functions of $\Omega_{rc}$ from various joint datasets for
 the DGP model.}
 \end{figure}
 \end{center}

The Dvali-Gabadadze-Porrati (DGP) model is a popular model
 which modifies the gravity to allow the cosmic acceleration
 without dark energy~\cite{r41,r42}. This model could arise
 from the braneworld theory in which gravity leaks out into
 the bulk on large scales. As is well known, for the flat
 DGP model (here we only consider the self-accelerating
 branch), the corresponding $E(z)$ is given by~\cite{r41,r42}
 \be{eq21}
 E(z)=\sqrt{\Omega_{m0}(1+z)^3+\Omega_{rc}}+
 \sqrt{\Omega_{rc}}\,,
 \ee
 where $\Omega_{rc}$ is a constant. It is easy to
 see that $E(z=0)=1$ requires
 \be{eq22}
 \Omega_{m0}=1-2\sqrt{\Omega_{rc}}\,.
 \ee
 Therefore, the DGP model has only one independent model
 parameter $\Omega_{rc}$. Notice that $0\leq\Omega_{rc}\leq 1/4$ is
 required by $0\leq\Omega_{m0}\leq 1$. It is easy to obtain the
 total $\chi^2$ as a function of the single model parameter
 $\Omega_{rc}$. In Fig.~\ref{fig8}, we plot the corresponding
 $\chi^2$ and likelihood ${\cal L}\propto e^{-\chi^2/2}$ from
 various joint datasets for the DGP model. In Table~\ref{tab5},
 we also present the $\chi_{min}^2$ and the best-fit model
 parameters (with $1\sigma$ and $2\sigma$ uncertainties) from
 various joint datasets for the DGP model. It is easy to see
 that Planck data favors a smaller $\Omega_{rc}$ than WMAP9
 data, while GRBs data favors a slightly smaller $\Omega_{rc}$.


 \begin{table}[b]
 \renewcommand{\arraystretch}{1.5}
 \begin{center}
 \begin{tabular}{l|ccc} \hline\hline
  &  $\chi^2_{min}$  & $\chi^2_{min}/dof$ &  $\Omega_{rc}$  \\ \hline
  SN+BAO+Planck\ \ & \ \ 665.697 \ \  & 1.146 & \ \ $0.1092_{-0.0038}^{+0.0038}\,
  (1\sigma)\,_{-0.0077}^{+0.0075}\,(2\sigma)$ \ \\ \hline
  SN+BAO+Planck+GRBs\ \ & \ \ 698.435 \ \  & 1.058 & \ \
   $0.1092_{-0.0038}^{+0.0038}\,(1\sigma)\,_{-0.0077}^{+0.0075}\,(2\sigma)$ \ \\\hline
  SN+BAO+WMAP9\ \ & \ \ 611.705 \ \  & 1.053 & \ \ $0.1339_{-0.0048}^{+0.0047}\,
  (1\sigma)\,_{-0.0096}^{+0.0092}\,(2\sigma)$ \ \\ \hline
  SN+BAO+WMAP9+GRBs~ \ \ & \ \ 644.750 \ \  & 0.977 & \ \ $0.1336_{-0.0047}^{+0.0046}\,
   (1\sigma)\,_{-0.0096}^{+0.0092}\,(2\sigma)$ \ \\
 \hline\hline
 \end{tabular}
 \end{center}
 \caption{\label{tab5} The $\chi_{min}^2$ and the best-fit
 model parameters (with $1\sigma$ and $2\sigma$
 uncertainties) from various joint datasets for the DGP model.}
 \end{table}



\subsection{SGCG model}\label{sec3e}

The Chaplygin gas (CG) model was firstly proposed
 by Kamenshchik {\it et al.}~\cite{r43}. In this model, the
 pressure $p$ of the fluid is related to its energy density
 $\rho$ through $p = -{\cal A}/\rho$, where $\cal A$ is a
 positive constant. In the literature, the generalized
 Chaplygin gas model (GCG)~\cite{r44} model is extensively
 considered, in which the equation of state for this fluid
 is generalized to
 \be{eq23}
 p=-{\cal A}/ \rho^{\eta}\,,
 \ee
 where $\eta$ is a constant. Originally, the CG or GCG models
 were considered as promising models united dark matter and
 dark energy~\cite{r43,r44}, since this fluid can mimic
 pressureless matter in the early time and cosmological
 constant in the late time. However, this possibility was
 excluded later (see e.g.~\cite{r53}). Thus, in the literature,
 the CG or GCG are usually considered as a candidate of dark
 energy only, coexisting with dark matter.

\begin{center}
 \begin{figure}[tb]
 \centering
 \includegraphics[width=1.0\textwidth]{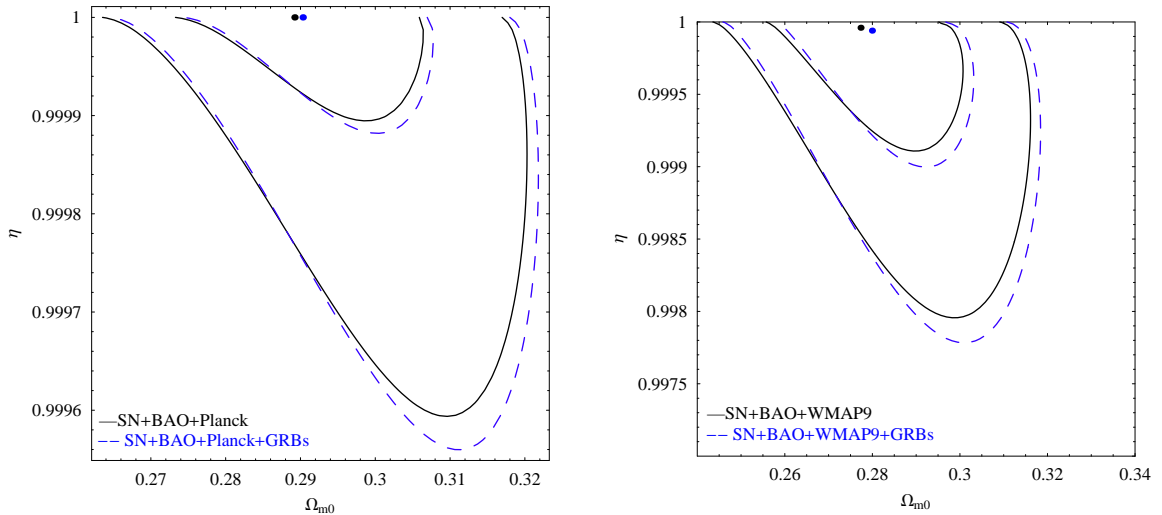}
 \caption{\label{fig9}
 The $68.3\%$ and $95.4\%$ confidence level contours in
 the $\Omega_{m0}-\eta$ parameter space from various joint datasets
 for the SGCG model. The best-fit parameters are indicated
 by a solid point.}
 \end{figure}
 \end{center}

In the GCG model, there are 3 free parameters, namely $\Omega_{m0}$,
 $\eta$ and $\cal A$. In~\cite{r45}, Lima {\it et al.}
 proposed a simplified GCG (SGCG) model, in which they argued that
 the parameter $\cal A$ could be related with $\eta$ according to
 ${\cal A}=\eta \rho_0^{1+\eta}$. So, the simplified equation
 of state becomes~\cite{r45,r46}
 \be{eq24}
 p=-\eta \rho_0 \left(\frac{\rho_0}{\rho}\right)^\eta\,.
 \ee
 As is argued in~\cite{r45}, $\eta>0$ is required by $p<0$ to
 accelerate the universe, while $\eta\leq 1$ is required by the
 causality, namely the adiabatic sound speed of this fluid
 cannot exceed the speed of light. So, as in
 e.g.~\cite{r45,r46,r47,r48}, we also restrict $0<\eta\leq 1$
 in this work. In the SGCG model, the corresponding $E(z)$ is
 given by~\cite{r45,r46,r47}
 \be{eq25}
 E(z)=\left\{\Omega_{m0}(1+z)^3+(1-\Omega_{m0})
 \left[(1-\eta)(1+z)^{3(1+\eta)}+\eta\right]^{1/(1+\eta)}
 \right\}^{1/2}\,.
 \ee
 Now, there are only two free parameters in this model, namely
 $\Omega_{m0}$ and $\eta$. By minimizing the corresponding
 total $\chi^2$, we find the best-fit parameters and present
 them in Table~\ref{tab6}. In Fig.~\ref{fig9}, we present the
 corresponding $68.3\%$ and $95.4\%$ confidence level contours
 in the $\Omega_{m0}-\eta$ parameter space from various joint
 datasets for the SGCG model. From Table~\ref{tab6} and
 Fig.~\ref{fig9}, it is easy to see that Plank data favors
 a larger $\Omega_{m0}$ and a slightly larger $\eta$ than WMAP9
 data, while GRBs data favors a slightly larger $\Omega_{m0}$
 and a slightly smaller $\eta$. However, it is obvious that all
 joint datasets favor $\eta\simeq 1$ (in this case the SGCG
 model reduces to the original CG model).


 \begin{table}[t]
 \renewcommand{\arraystretch}{1.5}
 \begin{center}
 \begin{tabular}{l|cccc} \hline\hline
  &  $\chi^2_{min}$  & $\chi^2_{min}/dof$ & $\ \Omega_{m0}$  & ~~~~~ $\eta$ ~~~~  \\ \hline
  SN+BAO+Planck\ \ & \ \ 563.352 \ \  & 0.971 & \ \ 0.2892 \ \ &\ \ 1.0      \ \\ \hline
  SN+BAO+Planck+GRBs \ \ & \ \ 597.135 \ \  & 1.068 & \ \ 0.2903 \ \ & \ \ 1.0     \ \\ \hline
  SN+BAO+WMAP9\ \ & \ \ 562.227 \ \  & 0.969 & \ \ 0.2773 \ \ &\ \ 0.99996 \ \\ \hline
  SN+BAO+WMAP9+GRBs \ \ & \ \ 596.233 \ \  & 0.905 & \ \ \ 0.2799 \ \ \ &\ \  0.99994  \ \\
  \hline\hline
 \end{tabular}
 \end{center}
 \caption{\label{tab6} The $\chi_{min}^2$ and the best-fit
 model parameters from various joint datasets for the SGCG
 model.}
 \end{table}



\vspace{-12mm} 


\subsection{RDE model}\label{sec3f}

\begin{center}
 \begin{figure}[tbp]
 \vspace{3mm} 
 \centering
 \includegraphics[width=1.0\textwidth]{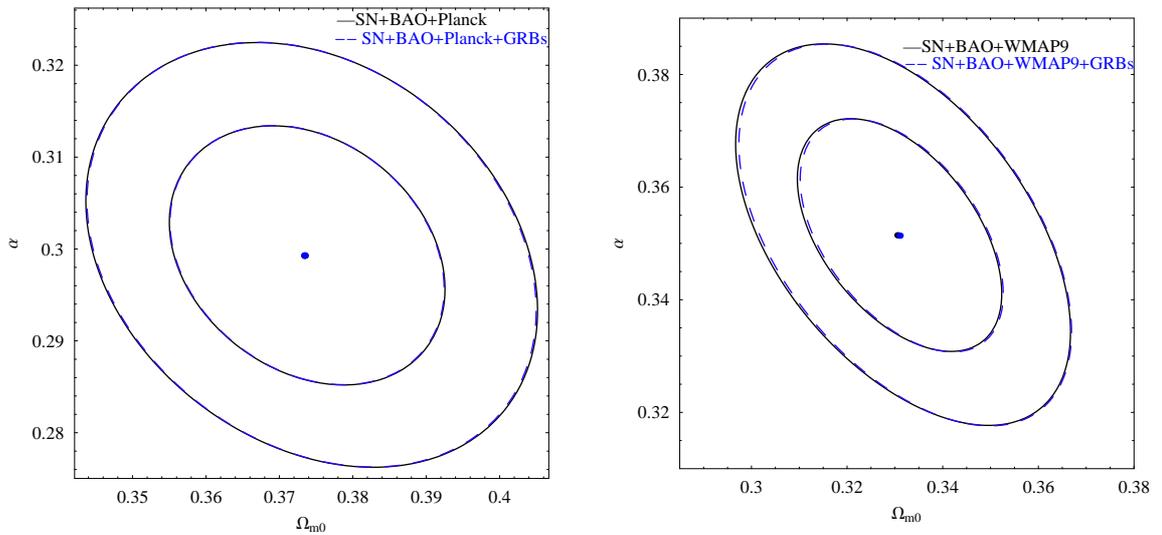}
 \caption{\label{fig10}
 The $68.3\%$ and $95.4\%$ confidence level contours in
 the $\Omega_{m0}-\alpha$ parameter space
 from various joint datasets for the RDE model. The best-fit
 parameters are indicated by a solid point.}
 \end{figure}
 \end{center}

The so-called holographic dark energy (HDE) has been studied
 extensively in the literature. Based on the holographic
 principle, it is argued that the density of dark energy
 is given by $\rho_{de}=3c^2 M_p^2 L^{-2}$, where $M_p$ is
 the reduced Planck mass; $c$ is a numerical constant
 characterizing some uncertainties in the effective quantum
 field theory, and $L$ is the IR cut-off. In~\cite{r49}, Gao
 {\it et al.} proposed the so-called Ricci dark energy (RDE)
 model, which can be regarded as a variant of HDE model. In
 this model, the IR cut-off $L$ is chosen to be proportional
 to the Ricci scalar curvature radius, and hence
 $L^{-2}\propto \dot{H}+2H^2$. So, the density of
 RDE reads~\cite{r49}
 \be{eq26}
 \rho_{de}=3\alpha M_p^2 \left(\dot{H}+2H^2\right)\,,
 \ee
 where $\alpha$ is a dimensionless constant. In this model, it
 is easy to find that~\cite{r49,r50}
 \be{eq27}
 E(z)=\left[\,\frac{2\Omega_{m0}}{\,2-\alpha}\,(1+z)^3+
 \left(1-\frac{2\Omega_{m0}}{\,2-\alpha}\right)
 (1+z)^{4-2/\alpha}\right]^{1/2}.
 \ee
 There are two free model parameters, namely $\Omega_{m0}$
 and $\alpha$. By minimizing the corresponding
 total $\chi^2$, we find the best-fit parameters and present
 them in Table~\ref{tab7}. In Fig.~\ref{fig10}, we present the
 corresponding $68.3\%$ and $95.4\%$ confidence level contours
 in the $\Omega_{m0}-\alpha$ parameter space from various joint
 datasets for the RDE model. From Table~\ref{tab7} and
 Fig.~\ref{fig10}, we see that Plank data favors a larger
 $\Omega_{m0}$ and a smaller $\alpha$, while GRBs data favors
 a slightly larger $\Omega_{m0}$. However, it is obvious that
 the effect of GRBs data on the constraints is fairly weak
 for the RDE model.


 \begin{table}[tbp]
 \renewcommand{\arraystretch}{1.5}
 \begin{center}
 \begin{tabular}{l|cccc} \hline\hline
  &  $\chi^2_{min}$  & $\chi^2_{min}/dof$ &  $\Omega_{m0}$  &  $\ \alpha$ \\ \hline
  SN+BAO+Planck\ \ & \ \ 615.549 \ \  & 1.061 & \ \ 0.3734 \ \ & \ \ 0.2993 \ \\ \hline
  SN+BAO+Planck+GRBs\ \ & \ \ 648.474 \ \  & 0.984 & \ \ 0.3734 \ \ & \ \ 0.2993 \ \\ \hline
  SN+BAO+WMAP9\ \ & \ \ 589.919 \ \  & 1.017 & \ \ 0.3305 \ \ & \ \ 0.3514 \ \\ \hline
  SN+BAO+WMAP9+GRBs~\ \ & \ \ 622.894 \ \  & 0.945 & \ \ \ \ 0.3310 \ \ \ \ & \ \ 0.3514 \ \\
 \hline\hline
 \end{tabular}
 \end{center}
 \caption{\label{tab7} The $\chi_{min}^2$ and the best-fit
 model parameters from various joint datasets for the RDE
 model.}
 \end{table}



\vspace{-10mm} 


\section{Conclusion and discussions}\label{sec4}

Gamma-ray bursts (GRBs) are among the most powerful sources
 in the universe. In the recent years, GRBs have been proposed
 as a complementary probe to type Ia supernovae (SNIa).
 However, as is well known, there is a circularity problem
 in the use of GRBs to study cosmology. In this work, based on
 the Pad\'e approximant, we propose a new cosmology-independent
 method to calibrate GRBs. We consider a sample consisting of
 138 long Swift GRBs and obtain 79 calibrated long GRBs at
 high-redshift $z>1.4$ (named Mayflower sample) which can be
 used to constrain cosmological models without the circularity
 problem. Then, we consider the constraints on several cosmological
 models with these 79 calibrated GRBs and other observational data.
 We show that GRBs are competent to be a complementary probe to
 the other well-established cosmological observations.

Some remarks are in order. First, in our
 calibration of GRBs, the Pad\'e approximant plays an important
 role. In fact, the present work is not the first one using the
 Pad\'e approximant in cosmology. We refer
 to e.g.~\cite{r54,r22inst,r57} for the previous relevant works. In
 these works, the Pad\'e approximant has been used in the
 slow-roll inflation, the reconstruction of the scalar field
 potential from SNIa, the data fitting of luminosity distance,
 the EoS parameterization, and the cosmological perturbation
 in LSS.

Second, when we calculate the errors for the distance
 moduli of the 59 low-redshift GRBs at $z_i<1.4$, the standard
 error propagation equation is used. In fact, there is an
 alternative way. Similar to e.g.~\cite{r27}, we can instead
 use the Monte Carlo method to evaluate the error propagations.
 That is, we generate a multivariate Gaussian distribution from
 the best-fit parameters and the corresponding covariance
 matrix. And then, we randomly sample $N$ suits (say, $N=10^6$)
 of the parameters $\{\alpha_0,\,\alpha_1,\,\alpha_2,\,
 \alpha_3,\,\beta_1,\,\beta_2\}$ from this distribution.
 For each suit of $\{\alpha_0,\,\alpha_1,\,\alpha_2,\,
 \alpha_3,\,\beta_1,\,\beta_2\}$, we can find the corresponding
 distance moduli of the 59 low-redshift GRBs from
 Eq.~(\ref{eq4}). After all, we can determine the means and the
 corresponding $1\sigma$ errors for the distance moduli of the
 59 low-redshift GRBs at $z_i<1.4$ from these $N$ samples. Of
 course, it is not surprising that the errors for the distance
 moduli of the 59 low-redshift GRBs obtained from the standard
 error propagation equation and the Monte Carlo method are
 coincident.

Third, we admit that the validity of the Amati
 relation is still in debate (we thank the referee for pointing
 out this issue). While many works support the Amati relation,
 it has also been seriously challenged in the literature (see
 e.g.~\cite{r61,r62}). In particular, it is
 argued in e.g.~\cite{r62} that there exists a significant
 problem in using the Amati Relation for cosmological purposes.
 Therefore, one should be careful and keep this issue in mind
 when using the Amati Relation in cosmology.

Fourth, the Mayflower sample of 79 calibrated GRBs obtained in
 the present work can be used to constrain cosmological models
 without the circularity problem. In this work, we have shown
 that GRBs are competent to be a complementary probe to the
 other well-established cosmological observations. However,
 as is shown in Sec.~\ref{sec3}, the inclusion of GRBs cannot
 considerably improve the constraints on most of
 the cosmological models (in fact, the constraints on some
 models become even worse). Therefore, to make GRBs into a
 competitively cosmological probe, one should accumulate more
 and more GRBs with smaller and smaller errors. We hope this
 could be done in the near future.

Finally, it is worth noting that when we fit the Union2.1 SNIa
 dataset~\cite{r24}, the numerical data given in~\cite{r63a}
 are used (see e.g. Eqs.~(\ref{eq3}), (\ref{eq5}),
 and Sec.~\ref{sec3}). Actually, this is equivalent to use the
 diagonal covariance matrix without systematics
 given in~\cite{r63b}. As is shown in e.g.~\cite{r24}, this
 diagonal covariance matrix without systematics might lead to
 an under-estimation of the errors (we thank the referee for
 pointing out this issue). To avoid this, we should use
 instead the full covariance matrix with systematics given
 in~\cite{r63c}. However, we choose not to redo all of the
 analysis with the full covariance matrix. As mentioned above,
 the inclusion of GRBs cannot considerably improve the
 constraints on most of the cosmological models (in fact,
 the constraints on some models become even worse), mainly due
 to the large errors of the current GRBs sample. If we redo
 all of the analysis with the full covariance matrix, this
 status cannot be changed and the constraints will become
 even worse, since the errors of GRBs coming from the full
 covariance matrix are larger. Noting that our main goal of
 this work is just to show that GRBs are competent to be a
 complementary probe to the other well-established cosmological
 observations, the conclusion will not be changed with using
 the full covariance matrix. In this sense, redoing all of the
 analysis with the full covariance matrix makes no significant
 difference. Nevertheless, we would like to remind the readers
 to be aware of this issue when using the Mayflower sample of
 79 calibrated GRBs in the relevant works.


\section*{ACKNOWLEDGEMENTS}
We thank the anonymous referee for quite useful comments and
 suggestions, which helped us to improve this work. We are grateful
 to Minzi~Feng, as well as Zu-Cheng~Chen, Xiao-Peng~Yan,
 Ya-Nan~Zhou, Xiao-Bo~Zou, Hong-Yu~Li, Shou-Long~Li, and
 Dong-Ze~Xue, for kind help and discussions. This work was
 supported in part by NSFC under Grants No.~11175016 and
 No.~10905005, as well as NCET under Grant No.~NCET-11-0790.

\newpage 

\renewcommand{\baselinestretch}{1.0}


\end{document}